\documentclass[10pt,journal]{IEEEtran}

\usepackage{graphicx}
\usepackage{psfrag}
\usepackage{subfigure}
\usepackage{amsmath,amssymb,amsfonts}
\usepackage{bm}
\usepackage{epsfig}
\usepackage{multirow}
\usepackage{algorithm}
\usepackage{algorithmic}
\usepackage{slashbox}
\usepackage{pict2e}
\usepackage{makecell}
\usepackage{dblfloatfix}
\DeclareMathOperator*{\argmax}{arg\,max\,} 

\DeclareMathOperator*{\argminA}{arg\,min\,}

\usepackage{hyperref} 
\hypersetup{hidelinks}

\title{Denoising Multiuser Interference Enabled Iterative AI Receiver}
\author{{\normalsize Xiangzhao Qin\,,~\IEEEmembership{Member,~IEEE}\,, Sha Hu\,, Jing Qian\,, Hao Wang}
\thanks{Xiangzhao Qin and Sha Hu are with Huawei Technologies Sweden AB, Lund, Sweden (e-mail: chrisqinxz@gmail.com, sha.hu@huawei.com), Jing Qian and Hao Wang are with Huawei Technologies Company, Ltd., Beijing, China (e-mail: qianjing3@huawei.com, hunter.wanghao@hisilicon.com).}
}

\begin{document}

\maketitle
\raggedbottom

\begin{abstract}
Finite-alphabet multiuser interference produces a non-Gaussian disturbance in multiple-input multiple-output (MIMO) receivers. Its covariance alone does not specify the likelihood required for Bayesian detection, limiting conventional minimum mean-square error interference rejection combining (MMSE-IRC). This article proposes an iterative artificial intelligence (AI) receiver that couples interference reconstruction, channel re-estimation, and desired-user detection through soft feedback. An interference-denoising network reconstructs the received interferer from residual observations, covariance information, and desired-symbol uncertainty. After cancellation, soft decisions yield least-squares channel estimates at data positions, which a channel network refines using the preceding estimate. MMSE-IRC preprocessing then supplies an effective channel and soft symbol estimates to a Transformer-ResNet-PIC detector. We relate these detector's inputs to the likelihood gradient of a relaxed Bayesian detection problem and develop two receiver architectures with stage-wise training. Link-level simulations show that learned interference cancellation improves channel estimation and coded detection, allowing the receiver to approach the relevant genie-interference reference in the evaluated multiuser scenarios. The proposed receiver substantially outperforms the conventional quasi-ML receiver based on LMMSE channel estimation and MMSE-IRC preprocessing.
\end{abstract}

\section{Introduction}
\IEEEPARstart{M}{ultiuser} multiple-input multiple-output (MIMO) transmission shares time--frequency resources among users and spatial streams. Reliable reception therefore depends on desired-channel estimation and interference modeling. Optimal multiuser detection exploits the users' alphabets and joint channel model \cite{Verdu1998Multiuser}. Conventional minimum mean-square error interference rejection combining (MMSE-IRC) instead suppresses interference through its covariance and is optimal among linear estimators with accurate second-order statistics \cite{Kay1993Estimation}.

We consider the system model $\bm{y}=\bm{H}\bm{s}+\bm{G}\bm{x}+\bm{n}$, where $\bm{s}$ and $\bm{x}$ are the desired and interfering symbols. For finite-alphabet $\bm{x}$ and Gaussian noise, $\bm{G}\bm{x}+\bm{n}$ generally follows a Gaussian-mixture distribution that its covariance does not fully characterize. Covariance-based filtering neither reconstructs the instantaneous interferer nor prevents its residual from contaminating data-aided channel estimates. The resulting errors propagate through soft feedback. We address these coupled errors by jointly refining interference reconstruction, channel estimation, and detection.

Existing receivers exchange soft information through interference cancellation and turbo MIMO processing \cite{Wang1999Turbo,Hochwald2003NearCapacity}. Learning has also been applied to physical-layer reception \cite{OShea2017Introduction}, joint channel estimation and detection \cite{Ye2018Power}, MIMO detection \cite{Samuel2019Learning,He2020ModelDriven}, soft interference cancellation \cite{Shlezinger2021DeepSIC}, and joint processing of pilot and data observations \cite{Honkala2021DeepRx}. These methods show how learned mappings can improve selected receiver operations. However, the interaction among instantaneous interference reconstruction, data-aided channel re-estimation, and desired-user detection has received less attention. In particular, uncertainty in the soft decisions should influence the next interference estimate, while the cleaned observation should in turn improve channel estimation and detection.

Following learned iterative inference \cite{Putzky2017RIM}, the proposed receiver first reconstructs $\bm{Gx}$ from residual observations, covariance information, and desired-symbol uncertainty. It then cancels this estimate and uses soft decisions to form data-position least-squares (LS) channel estimates. A channel network refines these estimates jointly with the preceding channel estimate. MMSE-IRC supplies an effective channel and linear soft symbols to a Transformer-ResNet-PIC detector based on attention \cite{Vaswani2017Attention}, residual learning \cite{He2016Residual}, and parallel interference cancellation (PIC). Updated symbol probabilities close the feedback loop. Appendix~\ref{Appendix_B} connects the detector inputs to a relaxed maximum a posteriori (MAP) likelihood gradient.

Sections II and III review related receivers and formulate the problem. Section IV presents the architecture, training, and complexity. Section V evaluates performance, and Section VI concludes. Appendices A and B provide the interference-estimation derivation and detector interpretation.

The notations in this article are as follows. Uppercase (lowercase) boldface letters denote matrices (column vectors). $(\cdot)^{H}$, $(\cdot)^{T}$, and $(\cdot)^{*}$ represent Hermitian transposition, transposition, and complex conjugation, respectively. $\|\cdot\|_{2}$ and $\|\cdot\|_{F}$ denote the Euclidean and Frobenius norms. $\mathbf{E}\{\cdot\}$ denotes expectation, $\mathcal{CN}(\bm{\mu},\bm{R})$ denotes a circularly symmetric complex Gaussian distribution with mean $\bm{\mu}$ and covariance $\bm{R}$, and $\bm{I}$ denotes an identity matrix of appropriate dimension. The space of $m\times n$ complex matrices is represented by $\mathbb{C}^{m\times n}$.

\section{Related Work on Iterative AI Receivers}
Classical iterative receivers combine soft cancellation with MMSE filtering \cite{Wang1999Turbo}, turbo equalization \cite{Tuchler2002Turbo}, and iterative MIMO detection/decoding \cite{Hochwald2003NearCapacity}. MMSE-PIC reduces detection complexity \cite{Studer2011MMSEPIC}; broader tradeoffs are reviewed in \cite{Yang2015MIMOSurvey}. Distribution-aware alternatives include partial marginalization \cite{Larsson2008PartialMarginalization}, prior-aided MMSE \cite{Tuchler2002MMSEAPriori}, GAMP \cite{Rangan2011GAMP}, OAMP \cite{Ma2017OAMP}, and robust non-Gaussian detection \cite{Wang1999Robust}. Their gains remain sensitive to likelihood and channel mismatch.

Learning-based reception encompasses end-to-end systems \cite{OShea2017Introduction,Ye2018Power}, model-free training \cite{AitAoudia2019ModelFree}, generative channel models \cite{Ye2020CGAN}, and trainable prototypes \cite{Cammerer2020Trainable}; hybrid analytical/learned processing is reviewed in \cite{Zappone2019WirelessDL}. Learned channel estimators address beamspace \cite{He2018BeamspaceCE} and time--frequency grids \cite{Soltani2019ChannelNet}, while ComNet combines estimation and detection \cite{Gao2018ComNet}. Detection methods include direct networks \cite{Samuel2017DeepMIMO}, unfolding \cite{Samuel2019Learning,He2020ModelDriven}, OAMP-Net \cite{He2018OAMPNet}, message passing \cite{Tan2020DNNMP}, MMNet \cite{Khani2020MMNet}, learned conjugate-gradient methods \cite{Wei2020LcgNet}, and hybrid symbol detection \cite{Farsad2021DataDriven}. ViterbiNet \cite{Shlezinger2020ViterbiNet}, DeepSIC \cite{Shlezinger2021DeepSIC}, and DeepRx \cite{Honkala2021DeepRx} learn selected receiver operations. Related trainable inference methods include recurrent machines \cite{Putzky2017RIM,Pratik2021REMIMO}, AMP \cite{Borgerding2017AMPNet}, sparse coding \cite{Gregor2010LISTA}, and belief-propagation decoding \cite{Nachmani2018Decoding}; see \cite{Monga2021Unrolling,Shlezinger2023ModelBased} for surveys.

Existing learned detectors do not necessarily reconstruct the instantaneous finite-alphabet interferer or refine the desired channel from the cleaned data observations. Our outer iteration connects these operations: interference reconstruction and data-aided channel estimation use the preceding soft decisions, and detection supplies updated probabilities for the next stage.

\section{Problem Formulation}
We first formulate interference cancellation for the considered MIMO system and derive the linear MMSE estimate of the received interference. We then describe how the interference-cancelled observation and preceding soft symbol decisions are used for iterative channel estimation and detection (ICED). This formulation identifies the quantities exchanged between receiver stages and shows how estimation errors can propagate across iterations.

\subsection{Interference Cancellation based Receiver Design}
Consider the complex-valued MIMO observation
\begin{equation}\label{eqn:mimo_system_model_intf}
\bm{y} = \bm{H} \bm{s} + \bm{G} \bm{x} + \bm{n},
\end{equation}
where $\bm{y}\in\mathbb{C}^{N_r\times1}$ is the received vector, $\bm{H},\bm{G}\in\mathbb{C}^{N_r\times N_t}$ are the desired- and interfering-user channel matrices and hold the same MIMO size, $\bm{s},\bm{x}\in\mathbb{C}^{N_t\times1}$ are their respective symbol vectors, and $\bm{n}\in\mathbb{C}^{N_r\times1}$ is additive white Gaussian noise (AWGN). Let $\bar{\bm{s}}\triangleq\mathbf{E}\{\bm{s}\}$ denote the soft symbol mean, $\Delta\bm{s}\triangleq\bm{s}-\bar{\bm{s}}$ its error, and $\bm{z}\triangleq\bm{G}\bm{x}$ the received interference. Subtracting the desired-user mean from \eqref{eqn:mimo_system_model_intf} gives
\begin{equation}\label{eqn:y_bar}
\bar{\bm{y}} \triangleq \bm{y} - \bm{H}\mathbf{E}\left\{\bm{s}\right\} = \bm{H} \Delta\bm{s} + \bm{z} + \bm{n}.
\end{equation}
For the linear MMSE (LMMSE) baseline, assume that $\bm{x}$ is zero mean and that $\bm{x}$, $\Delta\bm{s}$, and $\bm{n}$ are mutually uncorrelated. Their second-order moments are
\begin{equation*}
  \begin{aligned}
     & \mathbf{E}\left\{\bm{x}\bm{x}^{H}\right\}=\bm{I}, \\
     & \mathbf{E}\left\{\bm{n}\bm{n}^{H}\right\}=\sigma^{2}\bm{I}, \\
     & \bm{C}_{\bm{s}} \triangleq\mathbf{E}\left\{\Delta\bm{s}\Delta\bm{s}^{H}\right\}.
  \end{aligned}
\end{equation*}
We estimate the received interference directly by minimizing the mean squared error over linear maps
\begin{equation}
 \bm{W}_{\bm{z}}=\argminA_{\bm{W}_{\bm{z}}^{H}}\mathbf{E}\left\{\left\|\bm{z}-\bm{W}_{\bm{z}}^{H}\bar{\bm{y}}\right\|_{2}^{2}\right\}.
\end{equation}
The resulting estimate $\hat{\bm{z}}=\bm{W}_{\bm{z}}^{H}\bar{\bm{y}}$ is
\begin{equation}\label{eqn:Gx_est}
  \hat{\bm{z}} = \bm{G}\bm{G}^{H}\left(\bm{R}+\bm{H}\bm{C}_{\bm{s}}\bm{H}^{H}\right)^{-1}\bar{\bm{y}},
\end{equation}
by defining
\begin{equation*}
  \bm{R}\triangleq\bm{G}\bm{G}^{H}+\sigma^{2}\bm{I}.
\end{equation*}
For completeness, we provide a detailed derivation of \eqref{eqn:Gx_est} in Appendix ~\ref{Appendix_A}.
Under the stated assumptions, \eqref{eqn:Gx_est} is optimal among linear estimators. For finite-alphabet interference, it generally differs from the Bayesian conditional mean. An equivalent form is
\begin{equation}
  \hat{\bm{z}} = \left(\bm{R} - \sigma^{2}\bm{I}\right)\left(\bm{R}+\bm{H}\bm{C}_{\bm{s}}\bm{H}^{H}\right)^{-1}\bar{\bm{y}}.
\end{equation}
Subtracting $\hat{\bm{z}}$ from \eqref{eqn:mimo_system_model_intf} gives
\begin{equation}
\hat{\bm{y}} \triangleq \bm{y} - \hat{\bm{z}} = \bm{H}\bm{s} + \bm{G}\left(\bm{x} - \hat{\bm{x}}\right) + \bm{n},
\end{equation}
where $\hat{\bm{x}}\triangleq\bm{G}^{H}\left(\bm{R}+\bm{H}\bm{C}_{\bm{s}}\bm{H}^{H}\right)^{-1}\bar{\bm{y}}$ satisfies $\hat{\bm{z}}=\bm{G}\hat{\bm{x}}$.
The residual interference covariance used by the subsequent IRC stage is
\begin{equation}\label{eqn:G_est}
 \hat{\bm{R}} = \bm{G} \mathbf{E}\left\{(\bm{x}-\hat{\bm{x}})(\bm{x}-\hat{\bm{x}})^{H}\right\}\bm{G}^{H}+\sigma^{2}\bm{I}.
\end{equation}
The error covariance of the LMMSE symbol estimate is
\begin{equation} \label{eqn:x_mse}
\mathbf{E}\left\{(\bm{x}-\hat{\bm{x}})(\bm{x}-\hat{\bm{x}})^{H}\right\} = \bm{I} - \bm{G}^{H}\left(\bm{R}+\bm{H}\bm{C}_{\bm{s}}\bm{H}^{H}\right)^{-1}\bm{G}.
\end{equation}
Substituting \eqref{eqn:x_mse} into \eqref{eqn:G_est} yields
\begin{equation}\label{eqn:R_hat}
  \hat{\bm{R}} = \bm{R} -
  \left(\bm{R}-\sigma^{2}\bm{I}\right)\left(\bm{R}+\bm{H}\bm{C}_{\bm{s}}\bm{H}^{H}\right)^{-1}\left(\bm{R}-\sigma^{2}\bm{I}\right).
\end{equation}
When the desired-user soft decisions are exact, $\bm{C}_{\bm{s}}=\bm{0}$ and \eqref{eqn:R_hat} reduces to
\begin{equation}
\begin{split}
\hat{\bm{R}}
    &= \bm{G}\left(\bm{I}-\bm{G}^{H}\left(\bm{G}\bm{G}^{H}+\sigma^{2}\bm{I}\right)^{-1}\bm{G}\right)\bm{G}^{H} + \sigma^{2}\bm{I} \\
	&= \bm{G}\left( \mathrm{MMSE}(\bm{x})\right)\bm{G}^{H} + \sigma^{2}\bm{I}.
\end{split}
\end{equation}

\subsection{Iterative Channel Estimation and Signal Detection}
The iterative channel estimation and detection baseline uses successive interference cancellation (SIC) of the
desired-user streams
to obtain an LS channel estimate at each data position. For receive antenna $m$ and desired-user stream $n$, the update
at outer iteration $i$ is
\begin{equation} \label{eqn:LS_H}
\hat{H}_{ls,m,n}^{(i)} =  \left(\hat{y}_{m}^{(i)}-\sum\limits_{n'=1, n'\neq n}^{N_{t}}\hat{H}_{m,n'}^{(i-1)}\hat{s}_{n'}^{(i-1)}\right)/\hat{s}_{n}^{(i-1)},
\end{equation}
where $\hat{y}_{m}^{(i)} \triangleq \left(y_{m} -\hat{z}_{m}^{(i-1)} \right)$ is the observation after interference cancellation,
and $\hat H_{m,n'}^{(i-1)}$ and $\hat s_{n'}^{(i-1)}$ are the preceding channel and symbol estimates. Equation \eqref{eqn:LS_H} assumes $\hat s_n^{(i-1)}\ne0$, small-magnitude soft symbols require regularization in an implementation.

After re-estimating the channel, an LMMSE detector forms the next soft symbol estimate
\begin{equation}
\hat{\bm{s}}^{(i)} = \left({\hat{\bm{H}}}^{(i)} \right)^{H}\left(\hat{\bm{H}}^{(i)}\left(\hat{\bm{H}}^{(i)}\right)^{H}+\hat{\bm{R}}^{(i)}\right)^{-1}\hat{\bm{y}}^{(i)}.
\end{equation}
Here, $\hat{\bm{y}}^{(i)} \triangleq \left[\hat{y}_{1}^{(i)}, \hat{y}_{2}^{(i)}, \ldots, \hat{y}_{N_{r}}^{(i)}\right]^{T}$ and $\hat{\bm{R}}^{(i)}$ describes the residual disturbance.
The quality of $\hat{\bm{z}}^{(i)}$ depends on the preceding channel and soft symbol estimates. Its cancellation error then affects both $\hat{\bm{H}}^{(i)}$ and $\hat{\bm{s}}^{(i)}$. Moreover, the LMMSE updates retain only second-order information about finite-alphabet interference. These coupled errors motivate learned refinement of the interference, channel, and detection stages in the next section.

\section{Iterative AI Receiver With Interference Cancellation}
\begin{figure*}[!t]
    \centering
    \includegraphics[width=0.80\textwidth, keepaspectratio]{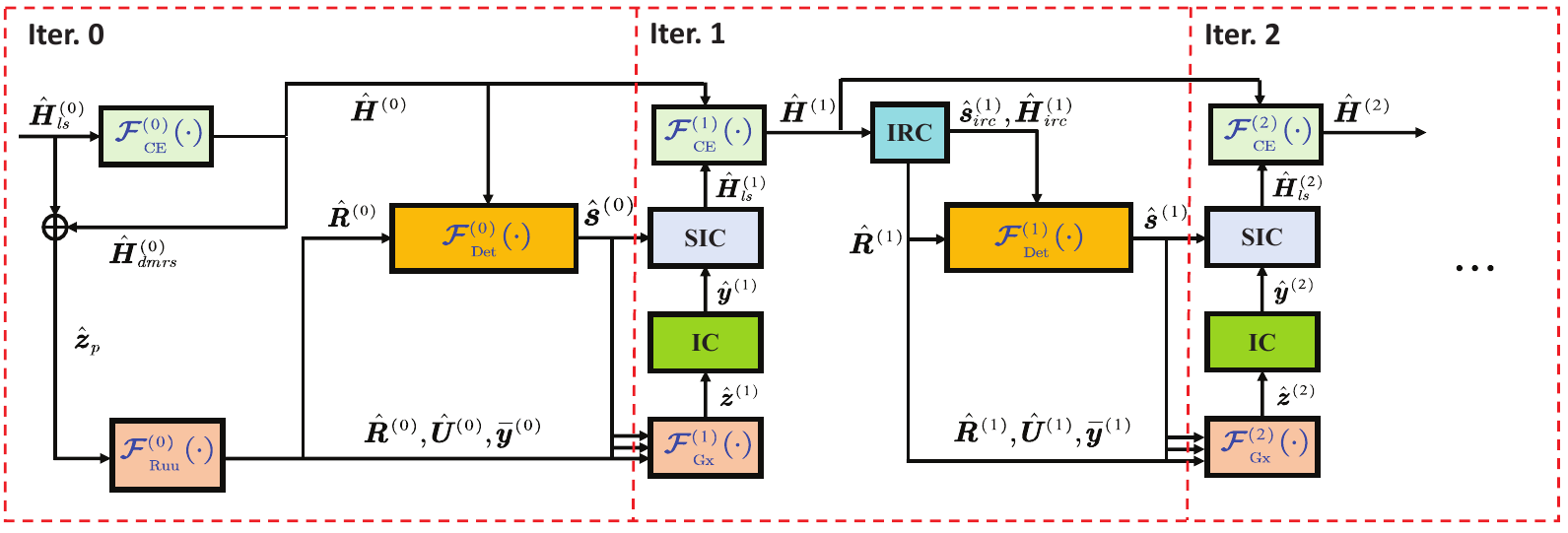}
    \caption{Block diagram of the proposed Type-I AI receiver.}
    \label{fig:Denoising_Iterative_AIReceiver_Arch_1}
\end{figure*}

\begin{figure*}[!t]
    \centering
    \includegraphics[width=0.80\textwidth, keepaspectratio]{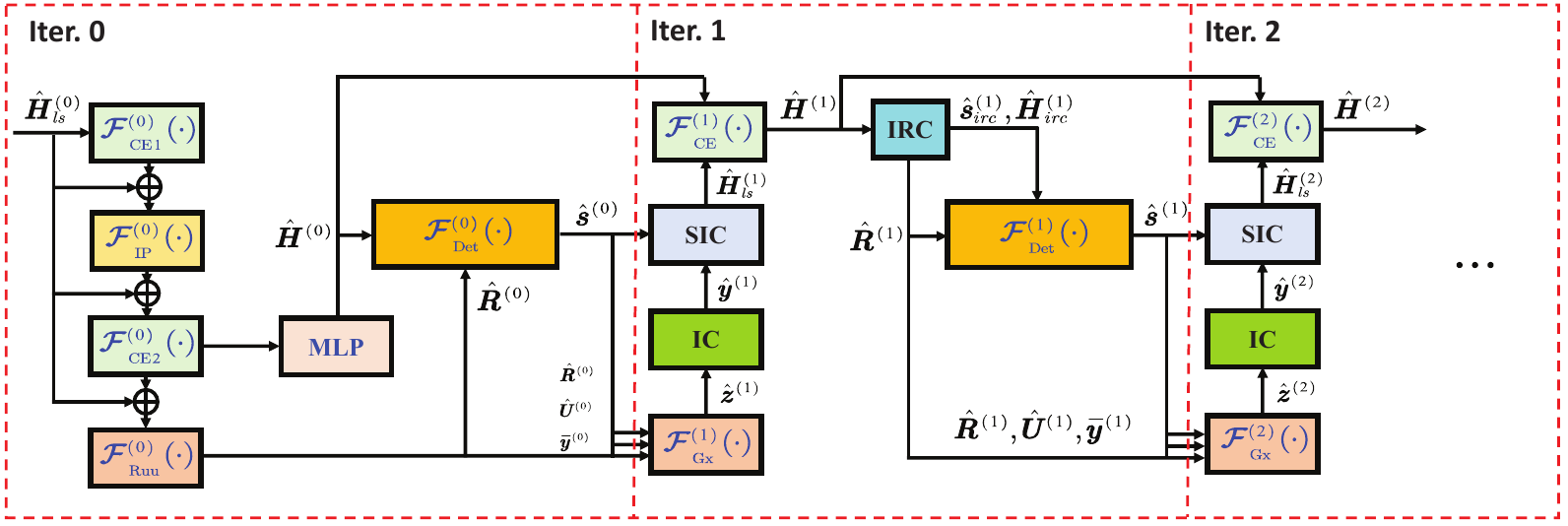}
    \caption{Block diagram of the proposed Type-II AI receiver.}
    \label{fig:Denoising_Iterative_AIReceiver_Arch_2}
\end{figure*}

Figures~\ref{fig:Denoising_Iterative_AIReceiver_Arch_1}
and ~\ref{fig:Denoising_Iterative_AIReceiver_Arch_2} illustrate two receiver
configurations that differ in their initial channel- and interference-estimation
paths. Both configurations subsequently exchange interference, channel, and symbol estimates
across outer iterations. We first describe the initialization shown for the Type-I receiver
and then the iterative operations common to the receiver design.

At iteration $i=0$ of Type-I receiver, the channel network
$\bm{\mathcal{F}}_{_{\mathrm{CE}}}^{(0)}(\cdot)$ takes the LS estimates at demodulation
reference signal (DMRS) positions, $\hat{\bm{H}}_{ls}^{(0)}\in\mathbb{C}^{N_p\times N_r\times N_t}$, and produces a data position estimate $\hat{\bm{H}}^{(0)} \in \mathbb{C}^{N_{d} \times N_{r}\times N_{t}}$. This estimate initializes detection and is retained for the next channel
estimation step. A second output $\hat{\bm{H}}_{dmrs}^{(0)} \in \mathbb{C}^{ N_{p} \times N_{r}\times N_{t}}$ provides a pilot position estimate. Its difference from $\hat{\bm{H}}_{ls}^{(0)}$ forms the covariance related feature
\begin{equation}\label{eqn:c_zp}
  \hat{\bm{C}}_{\bm{z}_{p}} \triangleq \left(\hat{\bm{H}}_{ls}^{(0)} - \hat{\bm{H}}_{dmrs}^{(0)}\right)\left(\hat{\bm{H}}_{ls}^{(0)} - \hat{\bm{H}}_{dmrs}^{(0)}\right)^{H},
\end{equation}
which the covariance network maps to the initial estimate $\hat{\bm{R}}^{(0)}\in\mathbb{C}^{N_d\times N_r\times N_t}$. The initial detector and first interference network use this estimate to start the feedback loop. The following subsections describe the operations for $i>0$.

\subsection{AI-Based Interference Cancellation}
The LMMSE estimate in \eqref{eqn:Gx_est} uses second-order statistics to predict $\bm{z}$.
To exploit the instantaneous observation, the interference network $\bm{\mathcal{F}}_{_{\mathrm{Gx}}}^{(i)}\left(\cdot\right)$ instead learns a mapping to
$\bm{z}$. Its input combines the preceding residual covariance, the desired-user uncertainty
mapped through the preceding channel estimate, and the residual observation
\begin{equation}
 \hat{\bm{U}}^{(i)} = \mathrm{concat}\left(\left[\hat{\bm{R}}^{(i-1)}, \hat{\bm{V}}^{(i-1)}, \bar{\bm{y}}^{(i-1)} \right]\right).
\end{equation}
Here, $\hat{\bm{V}}^{(i-1)}\triangleq \hat{\bm{H}}^{(i-1)}
 \hat{\bm{C}}_{\bm{s}}^{(i-1)}
\left(\hat{\bm{H}}^{(i-1)}\right)^{H}$ and $\bm{U}^{(i)} \in \mathbb{C}^{N_{d} \times N_{r} \times (2N_{r}+1)}$. The $n$-th
diagonal entry of the symbol error covariance is asymptotically calculated from the
detector's constellation probabilities
\begin{subequations}\label{eqn:symbol_moments}
\begin{align}
\mathrm{var}\left\{s_{n}^{(i-1)}\right\} &= \left(\sum\limits_{m=1}^{M}\alpha_{m}^{2} \cdot p\left(s_{n}^{(i-1)}=\alpha_{m}\right)\right) - \left(\hat{s}_{n}^{(i-1)}\right)^{2},  \\
\hat{s}_{n}^{(i-1)} &=  \left(\sum\limits_{m=1}^{M}\alpha_{m} \cdot p\left(s_{n}^{(i-1)}=\alpha_{m}\right)\right),
\end{align}
\end{subequations}
where $p\left(s_{n}^{(i-1)}=\alpha_{m}\right)$ is the detector probability assigned to constellation point $\alpha_{m}$.
These moments express how confidently the desired-user contribution can be removed. The network then produces
$\hat{\bm{z}}^{(i)} \in \mathbb{C}^{N_{d} \times N_{r}}$ as
\begin{equation}
  \hat{\bm{z}}^{(i)} = \bm{\mathcal{F}}_{_{\mathrm{Gx}}}^{(i)}\left(\hat{\bm{U}}^{(i)}\right).
\end{equation}
The estimate is passed to the channel estimation stage after substraction from the received signal.

\subsection{AI-Based Channel Re-Estimation}
The channel estimation stage starts with the interference cancellated observation
\begin{equation}
 \hat{\bm{y}}^{(i)} = \bm{y} - \hat{\bm{z}}^{(i)}.
\end{equation}
Using \eqref{eqn:LS_H}, the preceding soft symbols and $\hat{\bm{y}}^{(i)}$ provide data-position LS channel estimates
$\hat{\bm{H}}_{ls}^{(i)} \in \mathbb{C}^{N_{d} \times N_{r} \times N_{t}}$. Since these estimates inherit errors from
both the symbols and interference cancellation, the channel network also receives the preceding estimate $\hat{\bm{H}}^{(i-1)}$.
With $\tilde{\bm{H}}^{(i)}\triangleq\mathrm{concat}\left(\left[\hat{\bm{H}}_{ls}^{(i)},\hat{\bm{H}}^{(i-1)}\right]\right)$,
it produces
\begin{equation}
  \hat{\bm{H}}^{(i)} = \bm{\mathcal{F}}_{_{\mathrm{CE}}}^{(i)}\left(\tilde{\bm{H}}^{(i)}\right),
\end{equation}
where $\tilde{\bm{H}}^{(i)} \in \mathbb{C}^{N_{d} \times N_{r} \times 2N_{t}}$ and $\hat{\bm{H}}^{(i)} \in \mathbb{C}^{N_{d} \times N_{r} \times N_{t}}$.

\begin{figure}[!t]
    \centering
    \includegraphics[width=\columnwidth, keepaspectratio]{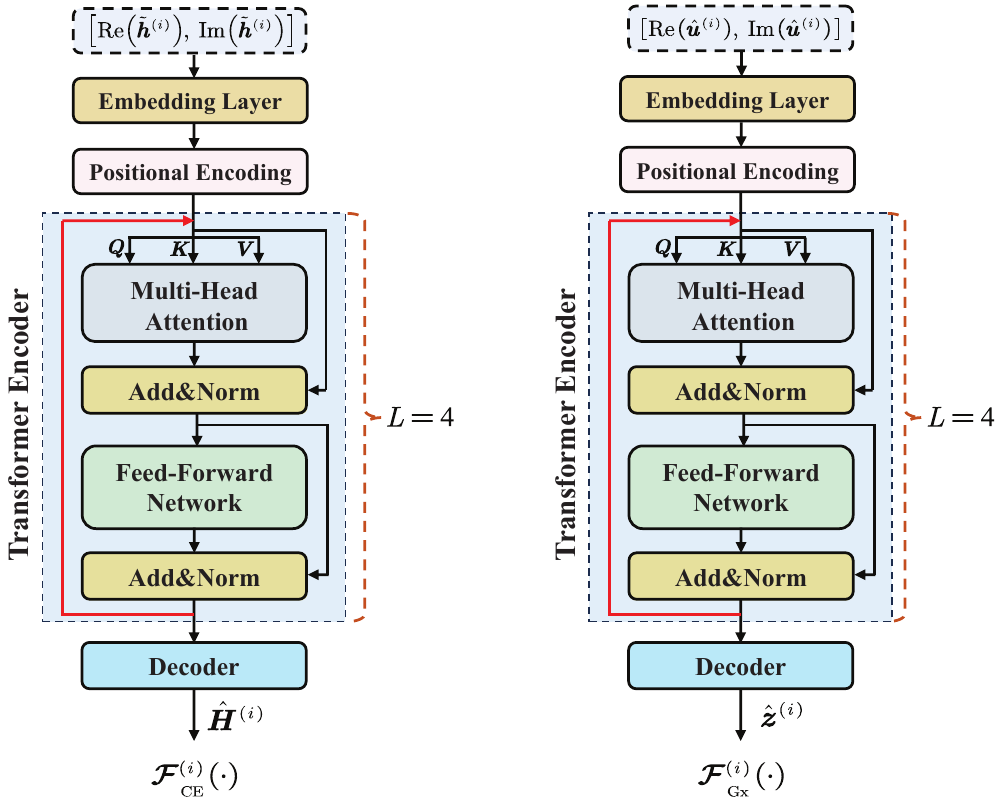}
    \caption{Network architectures for channel re-estimation and received interference reconstruction.}
    \label{fig:AI_CE_Gx}
\end{figure}

As shown in Fig.~\ref{fig:AI_CE_Gx}, both estimator networks contain an embedding layer, positional encoding,
and $L$ Transformer encoder layers. For the channel network, the real and imaginary parts of the concatenated estimates
from the input tokens
\begin{equation}
\begin{split}
\bar{\bm{h}}^{(i)}\triangleq {}&\mathrm{concat}\Big(
\big[\mathrm{Re}(\tilde{\bm{h}}^{(i)}),
\mathrm{Im}\left(\tilde{\bm{h}}^{(i)}\right)\big]\Big)\\
&\in\mathbb{R}^{N_{d}\times 4N_{r}N_{t}}.
\end{split}
\end{equation}
For the interference network, the corresponding tokens are
\begin{equation}
\begin{split}
\bar{\bm{u}}^{(i)}\triangleq {}&\mathrm{concat}\Big(
\big[\mathrm{Re}(\hat{\bm{u}}^{(i)}),
\mathrm{Im}(\hat{\bm{u}}^{(i)})\big]\Big)\\
&\in\mathbb{R}^{N_{d}\times 2N_{r}(2N_{r}+1)}.
\end{split}
\end{equation}
Here, $\tilde{\bm{h}}^{(i)} \triangleq \mathrm{vec}\left(\tilde{\bm{H}}^{(i)}\right) \in \mathbb{C}^{N_{d} \times 2N_{r}N_{t}}$
and $\hat{\bm{u}}^{(i)} \triangleq \mathrm{vec}\left(\hat{\bm{U}}^{(i)}\right) \in \mathbb{C}^{N_{d} \times N_{r}(2N_{r}+1)}$.
The two networks therefore share the same encoder structure while remaining task specific inputs and outputs.

\subsection{AI-Based Signal Detection}
\begin{figure*}[!t]
    \centering
    \includegraphics[width=0.98\textwidth, keepaspectratio]{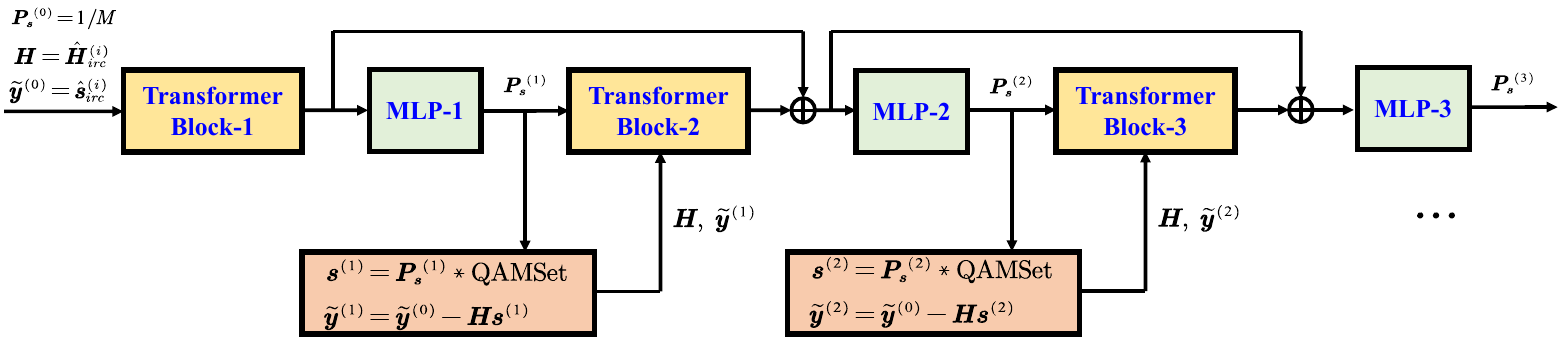}
    \caption{Transformer-ResNet-PIC architecture of $\bm{\mathcal{F}}_{_{\mathrm{Det}}}\left(\cdot\right)$ in the proposed AI receiver.}
    \label{fig:Transformer_ResNet_PIC}
\end{figure*}

After interference cancellation and channel re-estimation, MMSE-IRC maps the observation to two structured detector inputs
\begin{subequations}\label{eqn:mmse_irc}
\begin{align}
 \hat{\bm{H}}_{irc}^{(i)} &= \left(\hat{\bm{H}}^{(i)}\right)^{H} \left(\hat{\bm{H}}^{(i)}\left(\hat{\bm{H}}^{(i)}\right)^{H}+\hat{\bm{R}}^{(i)}\right)^{-1} \hat{\bm{H}}^{(i)}, \\
\hat{\bm{s}}_{irc}^{(i)} & = \left(\hat{\bm{H}}^{(i)}\right)^{H} \left(\hat{\bm{H}}^{(i)}\left(\hat{\bm{H}}^{(i)}\right)^{H}+\hat{\bm{R}}^{(i)}\right)^{-1}\hat{\bm{y}}^{(i)}.
\end{align}
\end{subequations}
The first quantity describes the effective coupling among desired-user streams, the second is the linear soft symbol estimate.
Both use the current channel estimate and residual disturbance covariance. For learned interference cancellation, the latter covariance represents the error remaining in the cleaned observation.
With $\bm{u}^{(i)} \triangleq \bm{z}-\hat{\bm{z}}^{(i)}+\bm{n}$, its target is
\begin{equation}
    \hat{\bm{R}}^{(i)} \approx \mathbf{E}\left\{\bm{u}^{(i)}\left(\bm{u}^{(i)}\right)^{H}\right\}.
\end{equation}
Unlike a sum of separate error and noise covariances, this expression also accounts for any correlation
introduced because $\hat{\bm{z}}^{(i)}$ is inferred from a noisy observation. The detector
$\bm{\mathcal{F}}_{_{\mathrm{Det}}}^{(i)}\left(\cdot\right)$ uses the two MMSE-IRC outputs
to update the desired-user symbols
\begin{equation}
\hat{\bm{s}}^{(i)} = \bm{\mathcal{F}}_{_{\mathrm{Det}}}^{(i)}\left(\hat{\bm{H}}_{irc}^{(i)}, \hat{\bm{s}}_{irc}^{(i)}\right).
\end{equation}
Its output supplies the soft decisions needed by the next interference and channel estimation steps.
Figure ~\ref{fig:Transformer_ResNet_PIC} shows the Transformer-ResNet-PIC implementation of this detector.

The detector comprises $L$ blocks, each containing a Transformer, a multilayer perceptron (MLP), and a PIC operation. At block $\ell$, the Transformer receives the effective channel $\bm{H} \in \mathbb{C}^{N_{d} \times N_{t} \times N_{t}}$,
the preceding residual $\tilde{\bm{y}}^{(\ell-1)} \in \mathbb{C}^{N_{d} \times N_{t} \times 1}$, and the preceding probabilities
$\bm{P}_{\bm{s}}^{(\ell-1)} \in \mathbb{R}^{N_{d} \times N_{t} \times M}$ by concatenating
\begin{equation}\label{eqn:trans_in}
\begin{split}
\bm{S}_{in}^{(\ell)}=\mathrm{concat}\Big(&\big[
\mathrm{Re}(\bm{H}),\mathrm{Im}(\bm{H}),\\
&\mathrm{Re}(\tilde{\bm{y}}^{(\ell-1)}),
\mathrm{Im}(\tilde{\bm{y}}^{(\ell-1)}),
\bm{P}_{\bm{s}}^{(\ell-1)}\big]\Big)\\
&\in\mathbb{R}^{N_d\times N_t\times L_{in}}
\end{split}
\end{equation}
with $L_{in} \triangleq 2(N_{t}+1)+M$. The effective channel and initial observation are initialized as
\begin{subequations}
  \begin{align}
    \bm{H} &= \hat{\bm{H}}_{irc}^{(i)}, \\
    \tilde{\bm{y}}^{(0)} &= \hat{\bm{s}}_{irc}^{(i)}.
  \end{align}
\end{subequations}
These quantities are the MMSE-IRC outputs defined in \eqref{eqn:mmse_irc}. The MLP maps the Transformer output $\bm{S}_{out}^{(\ell)} \in \mathbb{R}^{N_{d} \times N_{t} \times L_{out}}$ to constellation logits $\bm{L}_{\bm{s}}^{(\ell)} \in \mathbb{R}^{N_{d} \times N_{t} \times M}$, where $L_{out}$
is the Transformer embedding dimension. A softmax operation converts these logits to symbol probabilities
\begin{equation}\label{eqn:prob_softmax}
  \bm{P}_{\bm{s}}^{(\ell)} = \mathrm{softmax}\left(\bm{L}_{\bm{s}}^{(\ell)}\right) \in \mathbb{R}^{N_{d} \times N_{t} \times M}.
\end{equation}
These probabilities are fed to the next block. The PIC unit forms the soft symbol vector $\bm{s}^{\ell} \in \mathbb{C}^{N_{d} \times N_{t}}$ from their posterior mean
\begin{equation}\label{eqn:soft_decision}
  \bm{s}^{(\ell)} = \sum\limits_{m=1}^{M}\alpha_{m}\cdot\bm{P}_{\bm{s}}^{(\ell)}\left[:,:,m \right].
\end{equation}
It then updates the effective domain residual using the current soft decisions
\begin{equation}\label{eqn:pic}
  \tilde{\bm{y}}^{(\ell)} = \tilde{\bm{y}}^{(0)} - \bm{H}\bm{s}^{(\ell)}.
\end{equation}
Using $\tilde{\bm{y}}^{(0)}$ in every block prevents the combinations of earlier soft estimates from being subtracted repeatedly.
The Transformer encoder follows the structure illustrated for the estimator networks in Fig.~\ref{fig:AI_CE_Gx}. The
MLP and PIC operations are specific for detection. For completeness, Appendix~\ref{Appendix_B} provides the
mathematical foundation of the Transformer-ResNet-PIC detector.

\subsection{Training Pipeline of Iterative AI Receiver}
\begin{figure*}[!t]
    \centering
    \includegraphics[width=0.82\textwidth,keepaspectratio]{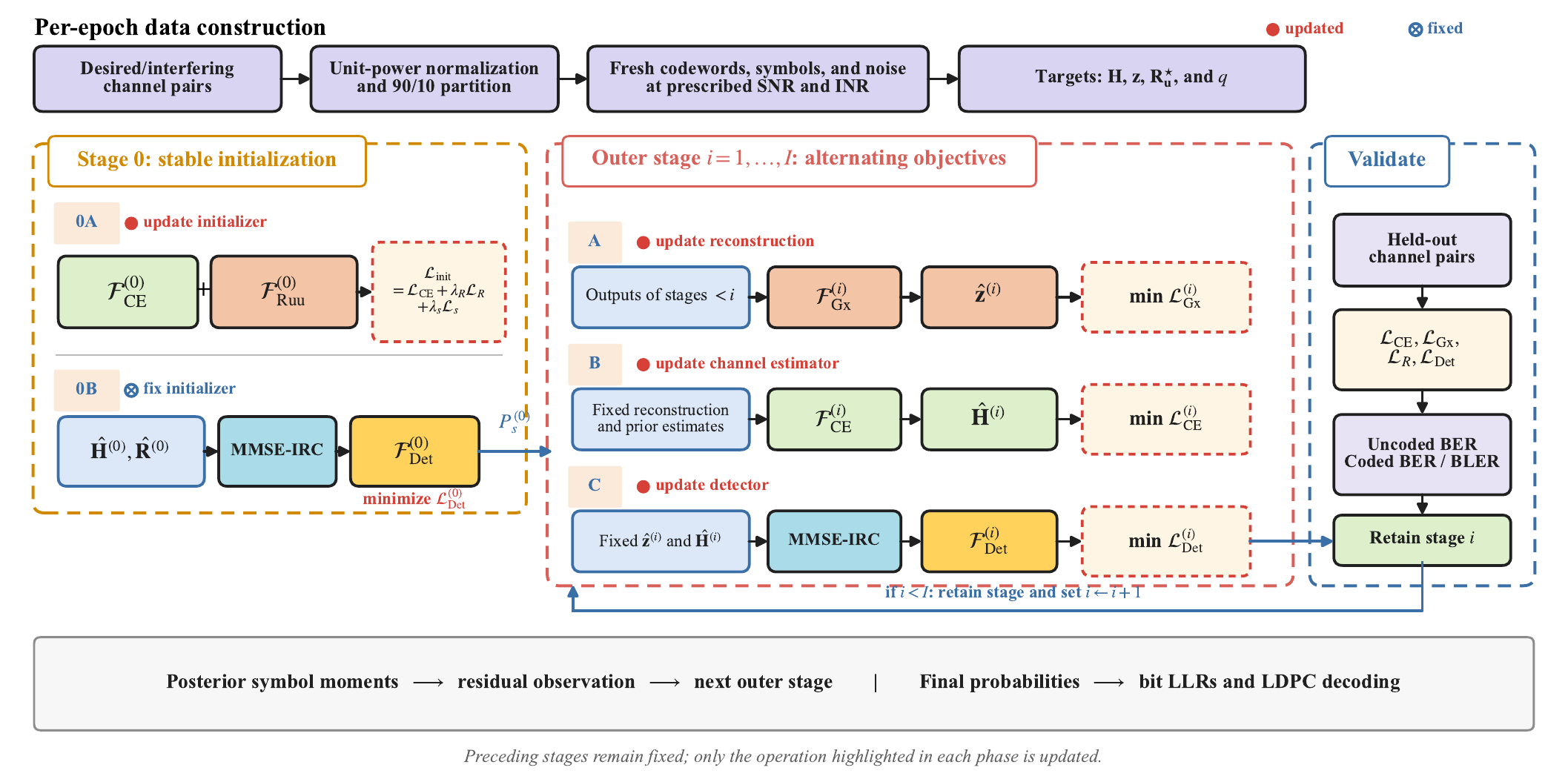}
    \caption{Stage-wise training pipeline of the iterative AI receiver. The initialization stage is trained in two phases, whereas each outer stage alternates interference reconstruction, channel re-estimation, and detection while retaining the preceding stages.}
    \label{fig:stagewise_training_pipeline}
\end{figure*}

We train greedily by freezing preceding stages and assigning a separate objective to each operation. Stable stage inputs limit error propagation, while alternating objectives avoid competition between MSE and cross-entropy losses. Independently paired, power-normalized channels are split $90\%/10\%$ for training/validation. Each epoch regenerates coded symbols and noise at the prescribed SNR and INR. Supervision comprises $\bm{H}$, $\bm{z}=\bm{Gx}$, $\bm{R}=\bm{GG}^{H}+\sigma^2\bm{I}$, and the transmitted constellation indices. Figure~\ref{fig:stagewise_training_pipeline} summarizes the schedule.

To match the real-valued network outputs with the complex-valued receiver quantities, we use the representation
\begin{equation}
 \rho(\bm{A})\triangleq
 \mathrm{concat}\Big(\big[\mathrm{Re}(\bm{A}),\mathrm{Im}(\bm{A})\big]\Big).
\end{equation}
Let $\langle\cdot\rangle_{\mathcal{B}}$ denote averaging over a mini-batch and all applicable data positions. At outer
stage $i$, we define the channel, interference, covariance, and detection losses as
\begin{subequations}\label{eqn:training_losses}
\begin{align}
\mathcal{L}_{\mathrm{CE}}^{(i)}
    &=\left\langle\left\|\rho\!\left(\hat{\bm{H}}^{(i)}\right)-\rho(\bm{H})\right\|_{F}^{2}\right\rangle_{\mathcal{B}},\\
\mathcal{L}_{\mathrm{Gx}}^{(i)}
    &=\left\langle\left\|\rho\!\left(\hat{\bm{z}}^{(i)}\right)-\rho(\bm{z})\right\|_{2}^{2}\right\rangle_{\mathcal{B}},\\
\mathcal{L}_{\mathrm{Ruu}}^{(i)}
    &=\left\langle\left\|\rho\!\left(\hat{\bm{R}}^{(i)}\right)-\rho\left(\bm{R}\right)\right\|_{F}^{2}\right\rangle_{\mathcal{B}},\\
\mathcal{L}_{\mathrm{Det}}^{(i)}
    &=-\frac{1}{L N_{t} N_{d}}\left\langle
      \sum_{\ell=1}^{L}\sum_{n=1}^{N_{t}}\sum_{k=1}^{N_{d}}
      \log P_{s,k,n}^{(i,\ell)}\left[q_{k,n}\right]
      \right\rangle_{\mathcal{B}},
\end{align}
\end{subequations}
where $q_{k,n}$ is the transmitted constellation index. The probability $P_{s,k,n}^{(i,\ell)}[q_{k,n}]$ is assigned by block $\ell$.
The MSE losses supervise the physical estimates; cross-entropy supervises every detector block so that each PIC update produces useful soft decisions. Initialization also uses the linear-symbol MSE
\begin{equation}
  \mathcal{L}_{\bm{s}}^{(0)} = \left\langle\left\|\rho\!\left(\hat{\bm{s}}^{(0)}\right)-\rho(\bm{s})\right\|_{F}^{2}\right\rangle_{\mathcal{B}}.
\end{equation}

We begin with the channel and covariance networks at $i=0$ and minimize
\begin{equation}\label{eqn:training_initial_loss}
  \mathcal{L}_{init} = \mathcal{L}_{\mathrm{CE}}^{(0)} + \lambda_{\mathrm{Ruu}}\mathcal{L}_{\mathrm{Ruu}}^{(0)}
  + \lambda_{\bm{s}}\mathcal{L}_{\bm{s}}^{(0)}.
\end{equation}
The two-stream weights are $(\lambda_{\mathrm{Ruu}},\lambda_{\bm{s}})=(0.4,0.2)$; the alternative configuration uses $(0.1,0)$. These weights prioritize channel accuracy while supervising covariance and initial symbols. We then freeze the initializer and train the initial detector with $\mathcal{L}_{\mathrm{Det}}^{(0)}$, providing stable channel and whitening inputs.

At stage $i>0$, frozen preceding stages provide the moments in \eqref{eqn:symbol_moments}. We average symbol variances over data positions and construct $\bar{\bm{y}}^{(i-1)}$ and $\hat{\bm{C}}_{\bm{s}}^{(i-1)}$. We train interference reconstruction, channel re-estimation, and detection successively:
\begin{equation}\label{eqn:training_alternating_objectives}
\begin{aligned}
  \textbf{Interference Phase:}\quad &\min_{\bm{\theta}_{\mathrm{Gx}}^{(i)}}\ \mathcal{L}_{\mathrm{Gx}}^{(i)},\\
  \textbf{Channel Phase:}\quad &\min_{\bm{\theta}_{\mathrm{CE}}^{(i)}}\ \mathcal{L}_{\mathrm{CE}}^{(i)},\\
  \textbf{Detection Phase:}\quad &\min_{\bm{\theta}_{\mathrm{Det}}^{(i)}}\ \mathcal{L}_{\mathrm{Det}}^{(i)},
\end{aligned}
\end{equation}
where the three $\bm{\theta}$ terms denote the active module parameters. Each phase freezes the other modules; $\mathcal{L}_{\mathrm{Ruu}}^{(i)}$ monitors residual-covariance accuracy, while $\mathcal{L}_{\mathrm{Gx}}^{(i)}$ drives interference learning. The completed stage initializes the next outer iteration. Algorithm~\ref{alg:iterative_ai_receiver_training} gives the procedure.

\begin{algorithm*}[!t]
\caption{Stage-Wise Training of the Iterative AI Receiver}
\label{alg:iterative_ai_receiver_training}
\normalsize
\begin{algorithmic}[1]
\REQUIRE Desired- and interfering-link channel sets, training SNR and INR, and number of outer stages $I$
\ENSURE Learned receiver parameters $\{\bm{\theta}^{(i)}\}_{i=0}^{I}$
\STATE Independently pair the two channel sets, normalize their average powers, and form the training and validation partitions.
\FOR{each initialization epoch}
    \STATE Generate coded symbols, received signals, and supervision targets.
    \STATE Update $\bm{\mathcal{F}}_{_{\mathrm{CE}}}^{(0)}\left(\cdot\right)$ and $\bm{\mathcal{F}}_{_{\mathrm{Ruu}}}^{(0)}\left(\cdot\right)$ using \eqref{eqn:training_initial_loss}.
\ENDFOR
\STATE Keep the initialization networks fixed and train $\bm{\mathcal{F}}_{_{\mathrm{Det}}}^{(0)}$ using $\mathcal{L}_{\mathrm{Det}}^{(0)}$.
\FOR{$i=1,\ldots,I$}
    \STATE Keep stages $0,\ldots,i-1$ fixed and obtain their channel, covariance, and symbol-probability estimates.
    \STATE Form the posterior symbol moments using \eqref{eqn:symbol_moments} and construct the residual observation for stage $i$.
    \STATE Train the interference-reconstruction operation using $\mathcal{L}_{\mathrm{Gx}}^{(i)}$.
    \STATE Hold the other operations fixed and train the channel-estimation operation using $\mathcal{L}_{\mathrm{CE}}^{(i)}$.
    \STATE Hold the estimator operations fixed and train the detection operation using $\mathcal{L}_{\mathrm{Det}}^{(i)}$.
    \STATE Evaluate the stage on the validation partition and retain its learned parameters.
\ENDFOR
\STATE Convert the final symbol probabilities to bit LLRs and evaluate uncoded BER, coded BER, and BLER.
\end{algorithmic}
\end{algorithm*}

Validation evaluates \eqref{eqn:training_losses} and converts the symbol probabilities to bit log-likelihood ratios (LLRs), followed by deinterleaving, rate recovery, and LDPC decoding. Uncoded BER, coded BER, and BLER reveal whether the stage-wise estimation gains improve decoded-packet reliability.

\subsection{Complexity Analysis}
We count the dominant online multiplications and divisions per OFDM slot, with $N_p$ pilot and $N_d$ data positions. Lower-order element-wise operations and probability-to-LLR conversion are omitted. LDPC decoding is common to all receivers and is excluded.

We first define the cost of a $K$-layer Transformer encoder that processes $N$ tokens, maps $p$ input features to an embedding of width $d$, uses a feed-forward width $f$, and produces $q$ output features as
\begin{equation}\label{eqn:transformer_complexity}
\begin{split}
\mathcal{C}_{\mathrm{T}}(N,p,q,d,f,K)
=\mathcal{O}\!\Big(&K(N^{2}d+Nd^{2}+Ndf)\\
&+Nd(p+d+q)\Big).
\end{split}
\end{equation}
The three terms inside the factor $K$ arise from attention-weight formation and application, the query/key/value and output projections, and the feed-forward network, respectively. The final term accounts for the embedding and readout mappings. Consequently, the received-interference and channel networks at an outer stage require
\begin{subequations}\label{eqn:estimator_complexities}
\begin{align}
\mathcal{C}_{\mathrm{Gx}}
&=\mathcal{C}_{\mathrm{T}}\!\left(
N_{d},2N_{r}(2N_{r}+1),2N_{r},d_{\mathrm{G}},f_{\mathrm{G}},K_{\mathrm{G}}
\right),\\
\mathcal{C}_{\mathrm{CE}}
&=\mathcal{C}_{\mathrm{T}}\!\left(
N_{d},4N_{r}N_{t},2N_{r}N_{t},d_{\mathrm{C}},f_{\mathrm{C}},K_{\mathrm{C}}
\right),
\end{align}
\end{subequations}
where $(d_{\mathrm{G}},f_{\mathrm{G}},K_{\mathrm{G}})$ and $(d_{\mathrm{C}},f_{\mathrm{C}},K_{\mathrm{C}})$ specify the two encoder configurations.
The initialization networks operate on the pilot positions. Including the learned interpolation from pilot to data positions, their costs can be written as
\begin{subequations}\label{eqn:initialization_complexities}
\begin{align}
\mathcal{C}_{\mathrm{CE}}^{(0)}
&=\mathcal{C}_{\mathrm{T}}\!\left(
N_{p},2N_{r}N_{t},2N_{r}N_{t},\right.\nonumber\\
&\hspace{17mm}\left.d_{0},f_{0},K_{0}
\right)
+\mathcal{O}(N_{r}N_{t}N_{p}N_{d}),\\
\mathcal{C}_{\mathrm{R}}^{(0)}
&=\mathcal{C}_{\mathrm{T}}\!\left(
N_{p},2N_{r}^2,2N_{r}^2,\right.\nonumber\\
&\hspace{17mm}\left.d_{\mathrm{R}},f_{\mathrm{R}},K_{\mathrm{R}}
\right)\nonumber\\
&\quad+\mathcal{O}\!\left(
N_{r}^2N_{p}N_{d}+(N_{p}+N_{d})N_{r}^3
\right).
\end{align}
\end{subequations}
The last term in \eqref{eqn:initialization_complexities} includes covariance interpolation and the eigenvalue, matrix-exponential, and Cholesky operations used to maintain a Hermitian positive-definite covariance estimate.

At each data position, MMSE-IRC forms an $N_r\times N_r$ matrix, factorizes or inverts it, and multiplies it by the channel and observation. Its cost is
\begin{equation}\label{eqn:mmse_irc_complexity}
\mathcal{C}_{\mathrm{IRC}}
=\mathcal{O}\!\left[
N_{d}\left(N_{r}^3+N_{r}^2N_{t}+N_{r}N_{t}^2\right)
\right].
\end{equation}
The residual-covariance update and the linear interference-reconstruction branch contain additional $N_{r}\times N_{r}$ products or inversions and therefore retain the same leading order. The SIC step in \eqref{eqn:LS_H} requires $\mathcal{O}(N_{d}N_{r}N_{t}^2)$ operations.

Let $L$ denote the number of Transformer-ResNet-PIC blocks, and let each block contain $K_{\mathrm{D}}$ Transformer encoder layers with embedding width $d_{\mathrm{D}}$ and feed-forward width $f_{\mathrm{D}}$. Processing the $N_{t}$ stream tokens independently at all $N_{d}$ data positions gives
\begin{equation}\label{eqn:detector_complexity}
\begin{split}
\mathcal{C}_{\mathrm{Det}}
=\mathcal{O}\!\Big(&LN_{d}\big[
K_{\mathrm{D}}(N_{t}^2d_{\mathrm{D}}+N_{t}d_{\mathrm{D}}^2
+N_{t}d_{\mathrm{D}}f_{\mathrm{D}})\\
&\qquad+N_{t}d_{\mathrm{D}}M+N_{r}N_{t}
\big]\Big),
\end{split}
\end{equation}
where the last two terms account for the constellation-logit readout and PIC update. Table~\ref{tab:complexity} summarizes the dominant orders by receiver stage, following the stage-wise presentation of the algorithm.

\begin{table*}[!t]
\centering
\normalsize
\caption{Computational Complexity of the Iterative AI Receiver}
\label{tab:complexity}
\renewcommand{\arraystretch}{1.0}
\begin{tabular*}{\textwidth}{@{\extracolsep{\fill}}ccl@{}}
\hline
Receiver stage & Operation & Dominant complexity \\
\hline
\multirow{3}{*}{Initialization}
& Channel and covariance estimation
& $\mathcal{C}_{\mathrm{CE}}^{(0)}+\mathcal{C}_{\mathrm{R}}^{(0)}$ \\
& MMSE-IRC preprocessing
& $\mathcal{C}_{\mathrm{IRC}}$ \\
& Transformer-ResNet-PIC detection
& $\mathcal{C}_{\mathrm{Det}}^{(0)}$ \\
\hline
\multirow{5}{*}{Outer stage $i$}
& Posterior moments and residual observation
& $\mathcal{O}\!\left[N_{d}N_{t}(M+N_{r})\right]$ \\
& Received-interference reconstruction
& $\mathcal{C}_{\mathrm{Gx}}^{(i)}$ \\
& SIC-LS and channel re-estimation
& $\mathcal{O}(N_{d}N_{r}N_{t}^2)+\mathcal{C}_{\mathrm{CE}}^{(i)}$ \\
& Residual covariance and MMSE-IRC
& $\mathcal{C}_{\mathrm{IRC}}^{(i)}$ \\
& Transformer-ResNet-PIC detection
& $\mathcal{C}_{\mathrm{Det}}^{(i)}$ \\
\hline
Total complexity
& \multicolumn{2}{c}{$\mathcal{C}_{\mathrm{AI}}$ in \eqref{eqn:total_ai_complexity}} \\
\hline
\end{tabular*}
\end{table*}

For $I$ outer stages after initialization, the total online complexity is
\begin{equation}\label{eqn:total_ai_complexity}
\begin{split}
\mathcal{C}_{\mathrm{AI}}
=&\ \mathcal{C}_{\mathrm{CE}}^{(0)}
+\mathcal{C}_{\mathrm{R}}^{(0)}
+\mathcal{C}_{\mathrm{IRC}}^{(0)}
+\mathcal{C}_{\mathrm{Det}}^{(0)}\\
&+\sum_{i=1}^{I}\Big[
\mathcal{C}_{\mathrm{Gx}}^{(i)}
+\mathcal{C}_{\mathrm{CE}}^{(i)}
+\mathcal{C}_{\mathrm{IRC}}^{(i)}
+\mathcal{C}_{\mathrm{Det}}^{(i)}\\
&\hspace{23mm}
+\mathcal{O}\!\left(N_{d}N_{t}M+N_{d}N_{r}N_{t}^2\right)
\Big].
\end{split}
\end{equation}
Thus, the receiver complexity grows linearly with the number of outer stages $I$. For large $N_{d}$, self-attention in the two estimator networks contributes the quadratic terms $\mathcal{O}(IN_{d}^2K_{\mathrm{G}}d_{\mathrm{G}})$ and $\mathcal{O}(IN_{d}^2K_{\mathrm{C}}d_{\mathrm{C}})$, whereas the detector grows linearly with $N_{d}$ because it applies attention only across the $N_{t}$ stream tokens at each data position. The matrix operations scale cubically with $N_{r}$, but $N_{r}$ is small in the considered MIMO configurations. In the implemented two-stream receiver, $N_{t}=2$, $N_{r}=4$, $M=16$, the initialization encoders use four layers with $d_{0}=f_{0}=128$, and the iterative channel estimator and detector use four-layer encoders with $d=512$ and $f=1024$; the detector contains $L=4$ PIC blocks. Hence, the dense Transformer projections constitute the principal cost for the evaluated dimensions.

Stage-wise training preserves this online order. Frozen stages perform forward inference, while only the active module stores gradients and executes its backward update. Training therefore adds a constant-factor backward cost for that module without retaining gradients through all preceding stages.

\section{Simulation Results}\label{sec:simulation_results}
Link-level simulations include coded-bit generation, DMRS insertion, soft detection, rate recovery, and LDPC decoding. We evaluate estimation errors and uncoded BER, coded BER, and BLER after initialization and each outer stage.

\subsection{Simulation Setup}\label{subsec:simulation_setup}
The desired and cochannel interfering links occupy the same OFDM resources, with two or four transmit streams and the corresponding receive dimensions. Each slot contains $14$ OFDM symbols and $24$ active subcarriers over two resource blocks at $15$ kHz spacing. Both users employ $16$-QAM; the interfering-user LDPC rate is $2/5$, and decoding uses at most $20$ iterations.

Type-A DMRS occupies OFDM symbols $3$ and $12$ (zero-based indices $2$ and $11$) with a two-subcarrier pilot spacing. This gives $N_{p}=24$ pilot REs per transmit port and $N_{d}=288$ data REs per stream after excluding the union of pilot positions. The sparse pilots motivate data-aided channel refinement.

The default channel uses the extended typical urban (ETU) profile, $70$ Hz maximum Doppler, and a $20$-ray Jakes realization with zero transmit and receive spatial correlation. We independently pair channels from two sets of $10{,}000$ realizations, normalize each link to unit average power, and reserve $18{,}000$ pairs for training and $2{,}000$ for validation.

Normalizing desired-signal power to unity gives noise power $10^{-\mathrm{SNR}/10}$ and interference power $10^{(\mathrm{INR}-\mathrm{SNR})/10}$. The resulting signal-to-interference-plus-noise ratio (SINR) is
\begin{equation}\label{eqn:simulation_sinr}
\mathrm{SINR}
=10\log_{10}\!\left(
\frac{10^{\mathrm{SNR}/10}}{1+10^{\mathrm{INR}/10}}
\right),
\end{equation}
where $\mathrm{SNR}$ and $\mathrm{INR}$ represent signal-to-noise ratio and interference-to-noise ratio, respectively.

\begin{table}[!t]
\caption{Default System Configuration for the Numerical Simulations}
\label{tab:simulation_configuration}
\centering
\normalsize
\renewcommand{\arraystretch}{1.0}
\setlength{\tabcolsep}{3pt}
\begin{tabular*}{\columnwidth}{@{\extracolsep{\fill}}lc@{}}
\hline\hline
Parameter & Value \\
\hline
Occupied PRBs $N_{\mathrm{RB}}$ & $2$ \\
Subcarriers/symbols $N_{\mathrm{sc}}$, $N_{\mathrm{sym}}$ & $24$, $14$ \\
Subcarrier spacing $\Delta f$ & $15$~kHz \\
Modulation alphabet $\mathcal{A}$ & $16$-QAM \\
Pilot/data REs $N_{p}$, $N_{d}$ & $24$, $288$ \\
Interferer code rate $R_{c,\mathrm{I}}$ & $2/5$ \\
LDPC iterations $I_{\mathrm{LDPC}}$ & $20$ \\
Channel profile / maximum Doppler & ETU / $70$~Hz \\
Jakes rays $N_{\mathrm{ray}}$ & $20$ \\
Spatial correlation $\rho_t$, $\rho_r$ & $0$, $0$ \\
Channel pairs $N_{\mathrm{tr}}$, $N_{\mathrm{val}}$ & $18{,}000$, $2{,}000$ \\
Mini-batch size $\mathcal{B}$ & $4$ \\
Adam learning rate & $10^{-4}$ \\
Adam $(\beta_1,\beta_2)$ & $(0.9,0.999)$ \\
Initialization / outer stages & $1/2$ \\
\hline\hline
\end{tabular*}
\end{table}

Each epoch regenerates coded bits, interleavers, symbols, and Gaussian noise for the fixed channel pairs. We use the mini-batch setting in Table~\ref{tab:simulation_configuration} and Adam with learning rate $10^{-4}$, $(\beta_1,\beta_2)=(0.9,0.999)$, $\epsilon=10^{-8}$, and no weight decay. The receiver has one initialization stage and two outer stages. The initializer and detector start from their $20$- and $100$-epoch checkpoints; each subsequent interference, channel, and detection training phase runs for $200$ epochs with preceding stages frozen.

Validation performs LLR conversion, deinterleaving, rate recovery, and LDPC decoding. We retain channel, interference, and covariance MSEs and symbol cross-entropy alongside the BER and BLER metrics. Table~\ref{tab:simulation_configuration} lists the defaults; experiment-specific changes are stated below.

\subsection{Performance Evaluation}\label{subsec:performance_evaluation}
We first consider first interference scenario, which uses a $4\times4$ desired link, a rank-one interfering link, an LDPC code rate of $R_c=2/3$, and $\mathrm{INR}=20$~dB. Both users employ $16$-QAM over two bundled physical resource blocks (PRBs), and the antenna branches are uncorrelated. For a given SNR, all receiver variants use the same held-out channels, transmitted packets, and noise realizations. This paired evaluation removes differences caused by Monte Carlo sampling and attributes the measured BLER separation to the receiver processing.
\begin{figure}[!t]
\centering
\includegraphics[width=\columnwidth,keepaspectratio]{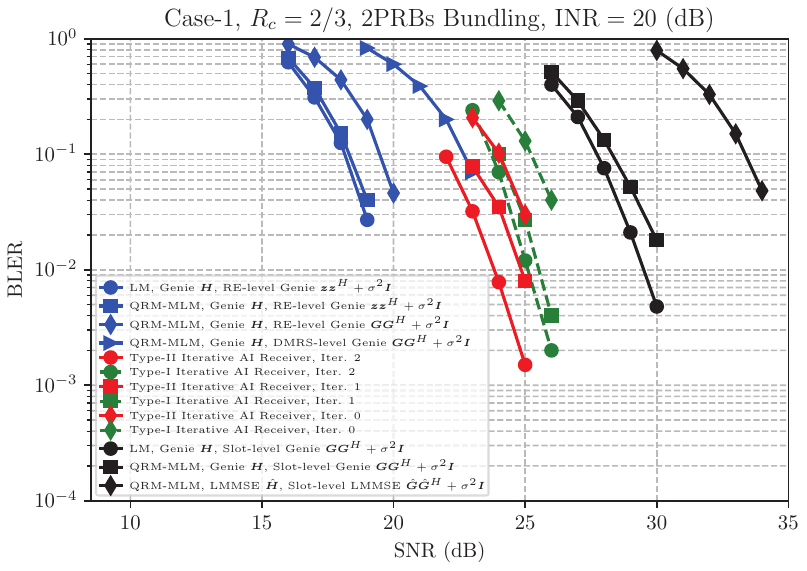}
\caption{BLER performance with $R_c=2/3$, $\mathrm{INR}=20$~dB, two-PRB bundling, a rank-four desired link, a rank-one interfering link, $16$-QAM, and $\rho_{t}=\rho_{r}=0$.}
\label{fig:bler_results_case_i}
\end{figure}

Figure~\ref{fig:bler_results_case_i} compares the proposed Type-I and Type-II receivers with several detector and interference-rejection combining (IRC) baselines. The log-MAP (LM) detector evaluates the complete log-domain symbol metric, whereas max-log-MAP (MLM) retains only its dominant term. QRM-MLM first prunes the QR-domain search tree and then evaluates the MLM metric over the surviving candidates; it lowers the search complexity at the expense of a possible detection loss. The genie-aided baselines use the true desired-user channel $\bm{H}$ and differ in the disturbance covariance supplied to the detector. The strongest reference uses the instantaneous, RE-level covariance $\bm{z}\bm{z}^{H}+\sigma^{2}\bm{I}$, where $\bm{z}=\bm{G}\bm{x}$. Replacing $\bm{z}\bm{z}^{H}$ by $\bm{G}\bm{G}^{H}$ removes the instantaneous finite-alphabet symbol information, while evaluating this covariance only at the DMRS positions or averaging it over an entire slot progressively reduces its time--frequency resolution. The practical conventional baseline combines an LMMSE estimate $\hat{\bm{H}}$, an LMMSE slot-level estimate of $\hat{\bm{G}}\hat{\bm{G}}^{H}+\sigma^{2}\bm{I}$, and QRM-MLM detection. These references separate the effects of tree-search approximation, channel-estimation error, and covariance mismatch.

The proposed receivers report the BLER after initialization ($i=0$) and after two interference-cancellation updates ($i=1$ and $i=2$). At every iteration, the receiver reconstructs the received interference, cancels it from the observation, refines the desired-user channel, and recomputes the soft symbol probabilities. Consequently, the separation between successive curves measures the benefit of the complete iterative update rather than a change in detector type.

Both architectures obtain a consistent gain from iteration. At $\mathrm{BLER}=10^{-1}$, the Type-II receiver requires approximately $24.0$ dB at $i=0$. After the first iteration, its BLER is already below $10^{-1}$ at the lowest evaluated point of $23$ dB, which places the corresponding SNR threshold below $23$ dB. The second iteration reaches the same target at approximately $22$ dB. The two interference cancellation iterations therefore reduce the required SNR by about $2$ dB. For the Type-I receiver, the SNR thresholds at $i=0$, $i=1$, and $i=2$ are approximately $25.2$, $24.0$, and $23.7$~dB, respectively. The first iteration provides a gain of about $1.2$ dB, whereas the second iteration contributes a further gain of approximately $0.3$ dB. The monotonic leftward shift of both sets of curves confirms that the stage-wise training procedure produces stable interference cancellation updates instead of amplifying reconstruction errors across stages.

The Type-II receiver also outperforms Type-I at every common iteration. At $\mathrm{BLER}=10^{-1}$, the final Type-II and Type-I stages require approximately $22.0$ and $23.7$~dB, respectively, corresponding to a Type-II advantage of about $1.7$~dB. Near $\mathrm{BLER}=10^{-2}$, their required SNRs are approximately $23.8$ and $25.1$~dB. This persistent separation shows that the receiver initialization materially affects the quality of the subsequent interference reconstruction, i.e., a more informative initial state gives each following cancellation and channel-refinement stage a cleaner observation.

The conventional references further clarify the source of the gain. The LM detector with genie $\bm{H}$ and RE-level genie $\bm{z}\bm{z}^{H}+\sigma^{2}\bm{I}$ achieves $\mathrm{BLER}=10^{-1}$ at approximately $18.2$~dB. Because this reference assumes exact knowledge of both the desired-user channel and the instantaneous interference covariance at every RE, quantities unavailable to a practical receiver, thus, it represents an unattainable genie-aided upper bound on receiver performance. Replacing the instantaneous covariance by RE-level $\bm{G}\bm{G}^{H}+\sigma^{2}\bm{I}$ increases this threshold to about $19.5$~dB, and using its DMRS-level counterpart raises it further to approximately $22.7$~dB. The proposed Type-II receiver at $i=2$ reaches the same BLER at about $22.0$~dB and therefore slightly surpasses the DMRS-level genie-covariance reference, despite operating without genie channel or interference information. This result demonstrates that reconstructing the instantaneous received interferer is more effective than relying on a covariance that is accurate only at pilot positions.

With slot-level genie covariance, LM and QRM-MLM require approximately $27.7$ and $28.3$~dB, respectively, at $\mathrm{BLER}=10^{-1}$. The practical baseline with LMMSE $\hat{\bm{H}}$ and LMMSE slot-level covariance requires about $33.4$~dB. The final Type-II and Type-I receivers therefore provide gains of approximately $11.4$ and $9.7$~dB, respectively, over this practical baseline. Moreover, the practical baseline remains above $4\times10^{-2}$ at 34~dB, whereas both proposed receivers reach BLERs on the order of $10^{-3}$ within the evaluated SNR range. The large gap shows that improving the detector alone cannot overcome a coarse interference model. The principal gain comes from learning the structured, RE-dependent interference, cancelling it before channel refinement, and propagating the improved channel and soft-symbol information to the next receiver stage.

\begin{figure}[!t]
\centering
\includegraphics[width=\columnwidth,keepaspectratio]{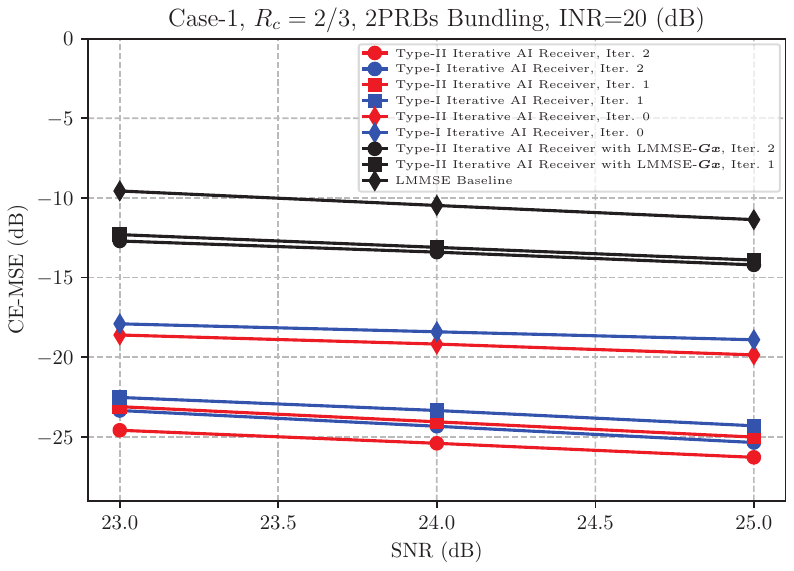}
\caption{Channel-estimation MSE in Case~1 under the same configuration as Fig.~\ref{fig:bler_results_case_i}.}
\label{fig:mse_results_case_i}
\end{figure}

Figure~\ref{fig:mse_results_case_i} demonstrates that the proposed Type-I and Type-II iterative AI Receivers progressively
recover channel estimation accuracy through the joint use of interference cancellation for interfering-user and desired-user,
and data-aided channel re-estimation. At each outer iteration, the $\bm{\mathcal{F}}_{_{\mathrm{Gx}}}^{(i)}$ network
first reconstructs and subtracts the instantaneous received interferer. The receiver then uses the preceding
soft symbol decisions and the cleaned data-bearing observations to form new channel estimates at the data REs,
and the $\bm{\mathcal{F}}_{_{\mathrm{CE}}}^{(i)}$ network fuses these data-aided channel estimates with the
preceding estimate. The proposed update therefore converts reliably detected REs into additional channel observations
beyond the sparse DMRS. Equally importantly, removing the interferer before this data-aided update prevents structured
interference from being absorbed into the desired-user channel estimate.

The proposed receivers obtain a large gain from this joint operation. At $\mathrm{SNR}=25$ dB, the Type-I CE-MSE from
$-18.9$ dB at initialization to $-24.3$ dB after the first iteration and $-25.4$ dB after the second update. Type-II
improves from $-19.8$ dB to $-25.0$ dB and $-26.3$ dB over the same stages. Hence, interference cancellation followed
by data-position channel re-estimation reduces the CE-MSE by approximately $6.4$ dB for both architectures relative
to their initial stages. The second update provides a smaller but still positive gain, showing that both $\bm{\mathcal{F}}_{_{\mathrm{Gx}}}^{(i)}$-related receivers converge toward stable channel estimates rather than
amplifying feedback errors. At the finial stage, Type-II maintains an approximately $0.9$-$1.2$ dB advantage over Type-I,
while both receivers outperforms the DMRS domain LMMSE baseline by $14-15$ dB over the evaluated SNR range.

The LMMSE-$\bm{Gx}$ control isolates the importance of learned interference reconstruction. This receiver replaces
only $\bm{\mathcal{F}}_{_{\mathrm{Gx}}}^{(i)}$ with the analytical estimator in \eqref{eqn:Gx_est}, while remaining
the subsequent data-position channel re-estimation. At $\mathrm{SNR}=25$ dB, its first and second updates achieve
CE-MSEs of only $-13.9$ dB and $-14.2$ dB, respectively, compared with $-11.4$ dB for the LMMSE baseline.
Thus, the first iteration provides a modest $2.5$ dB gain, and the second provides only another $0.3$ dB gain.
The same early saturation appears over SNR range from $23$ to $25$ dB, and the final LMMSE-$\bm{Gx}$ control
remains approximately $12$ dB worse than the final $\bm{\mathcal{F}}_{_{\mathrm{Gx}}}^{(i)}$ learned approach
for Type-II receivers. The limitation follows from the linear, second-order model underlying \eqref{eqn:Gx_est},
i.e., finite-alphabet interference, channel estimation error, and uncertain soft decisions, make the conditional
interference estimate nonlinear. Consequently, LMMSE-$\bm{Gx}$ leaves structured reconstruction errors in the
cleaned data observations. Treating there contaminated data REs as additional channel evidence restricts the
subsequent channel update and feeds part of the reconstruction error back to the next iteration, so additional
stages yield little improvement.

The recovered channel accuracy of the proposed receivers improves the subsequent detection and demodulation
stages. A more accurate $\hat{\bm{H}}$ yields better effective channel and soft symbol inputs for MMSE-IRC
and the Transformer-ResNet-PIC detector, thereby producing more reliable symbol probabilities and bit log-likelihood
ratios. These outputs improve soft demodulation and channel decoding and, when fed back, further strengthen the
next interference cancellation and data-aided channel re-estimation stage. The matching iteration wise trends in
Figs.~\ref{fig:bler_results_case_i} and~\ref{fig:mse_results_case_i} therefore show that the Type-I and Type-II
receivers obtain their performance gain through two coupled effects, i.e., direct suppression of the received
interferer and restoration of the desired-user channel estimate from cleaned data observations. The LMMSE-$\bm{Gx}$
control confirms that data-position channel re-estimation alone is insufficient when the preceding interference
reconstruction  remains mismatched. Accurate learned cancellation is required to unlock the full channel estimation,
detection, and demodulation gains.

We next consider the second interference scenario, in which both the desired and interfering links have rank two and their DMRS patterns collide. The desired user employs an LDPC code rate of $R_c=1/3$, both users transmit $16$-QAM symbols over two bundled PRBs, and the INR is fixed at $10$ dB. We use a low-correlation channel for both links. For every SNR, all receivers again process the same held-out channels, transmitted packets, and noise realizations. The colliding pilots make this setting particularly challenging because the receiver must separate two spatially multiplexed links from observations that contain their superposed DMRS signals.
\begin{figure}[!t]
\centering
\includegraphics[width=\columnwidth,keepaspectratio]{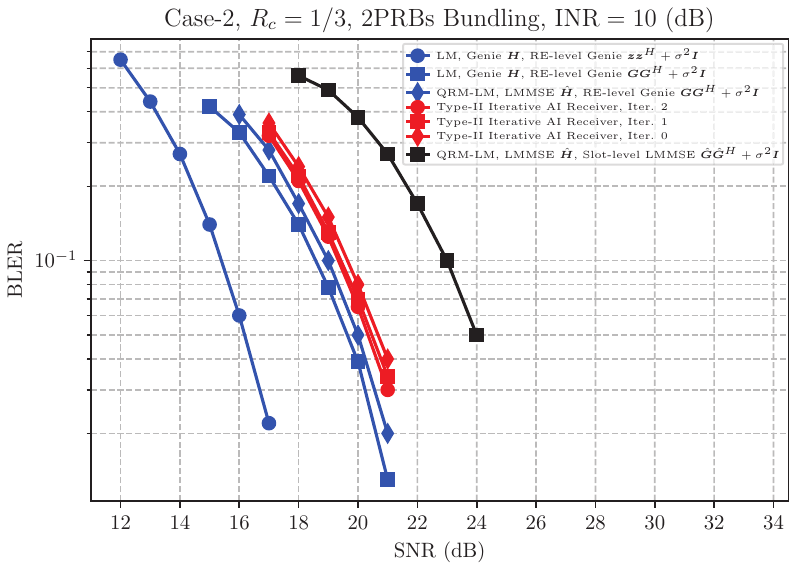}
\caption{BLER performance with $R_c=1/3$, $\mathrm{INR}=10$~dB, two-PRB bundling, rank-two desired and interfering links, colliding DMRS patterns, $16$-QAM, and $\rho_{t}=\rho_{r}=0$.}
\label{fig:bler_results_case_ii}
\end{figure}

Figure~\ref{fig:bler_results_case_ii} first shows that the Type-II iterative AI receiver has nearly reached its performance ceiling in this scenario. At $\mathrm{BLER}=10^{-1}$, Iter.~0 requires approximately $\mathrm{SNR}=19.6$ dB, while Iter.~1 and Iter.~2 reduce the threshold to approximately $19.4$ and $19.3$ dB, respectively. At $\mathrm{SNR}=21$ dB, the corresponding BLER values are $4.0\times10^{-2}$, $3.4\times10^{-2}$, and $3.0\times10^{-2}$. The monotonic but diminishing improvement from Iter.~0 to Iter.~2 demonstrates stable convergence and indicates that additional iterations can provide only limited gain. More importantly, the final receiver operates within approximately $0.3$ dB of the blue-diamond reference, which combines QRM-LM detection, LMMSE channel estimation, and genie RE-level interference information. Because this reference preserves the adopted detector and channel estimator while removing interference-information error, it defines the relevant theoretical performance ceiling for the proposed receiver. The Iter.~2 curve therefore effectively reaches this upper bound without access to genie interference information.

The blue-diamond and black-square references isolate the effect of interference knowledge. Both use QRM-LM detection and the same LMMSE channel estimate; they differ only in the interference information supplied to the detector. The blue-diamond reference uses genie RE-level interference information and reaches BLER targets of $10^{-1}$ and $5\times10^{-2}$ at approximately $\mathrm{SNR}=19$ and $\mathrm{SNR}=20$ dB, respectively. Replacing it with the practical slot-level interference estimate, as represented by the black-square curve, raises the corresponding SNR requirements to approximately $23$ and $24$ dB. This nearly $4$-dB penalty shows that interference estimation and reconstruction, rather than the detector or the desired-channel estimator, constitute the dominant performance bottleneck in this case. By learning and cancelling the instantaneous received interference at each RE, the proposed iterative AI receiver removes most of this penalty and approaches the genie-interference ceiling after only two updates.

We finally evaluate the Transformer-ResNet-PIC detector separately from the interference-reconstruction and channel-estimation networks. We exclude intercell interference and provide genie desired-user channel information so that the comparison isolates the detection operation.
The references are the LM, QRM-LM, and QRM-MLM detectors introduced above.
\begin{figure}[!t]
\centering
\includegraphics[width=\columnwidth,keepaspectratio]{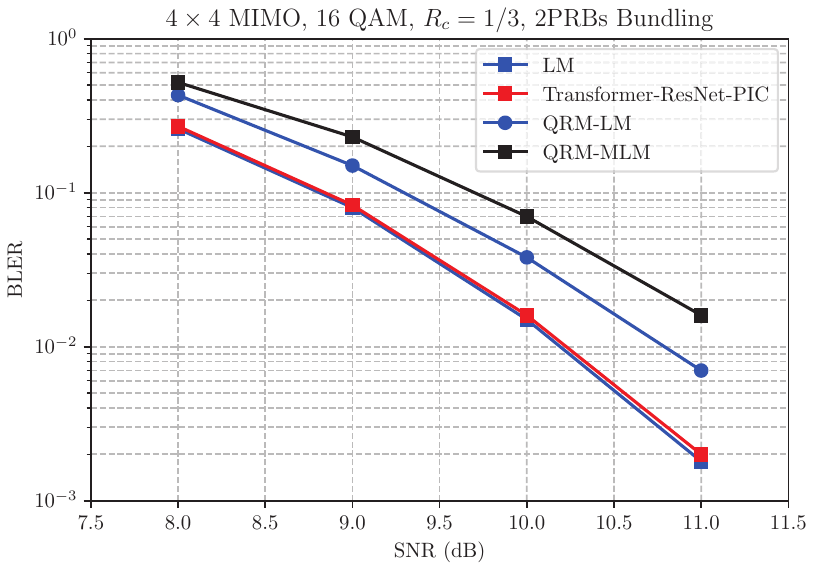}
\caption{Standalone detector BLER with $4\times4$ MIMO, $16$-QAM, $R_c=1/3$, an EVA channel, and $\rho_t=\rho_r=0.3$. Both detector tests use genie channels and no intercell interference.}
\label{fig:bler_transformer_resnet_pic_4x4mimo_16qam_rc13}
\end{figure}

Figure~\ref{fig:bler_transformer_resnet_pic_4x4mimo_16qam_rc13} considers a $4\times4$ MIMO link with $16$-QAM, $R_c=1/3$, and moderate spatial correlation. The Transformer-ResNet-PIC and LM curves nearly coincide over the evaluated SNR range.
The learned detector therefore closely approaches the genie-channel LM reference in this configuration and outperforms both reduced-search baselines.
This result provides a useful reference for interpreting the complete receiver. When accurate channel information is available and intercell interference is absent, the detector leaves little performance gap to LM in the low-rate configuration. Together with Figs.~\ref{fig:bler_results_case_i}--\ref{fig:bler_results_case_ii}, this observation supports the role of learned $\bm{Gx}$ reconstruction and data-aided channel re-estimation in the receiver gains, i.e., these operations improve the observation
and channel information supplied to the detector. The near coincidence in Fig.~\ref{fig:bler_transformer_resnet_pic_4x4mimo_16qam_rc13}, however, applies to the evaluated configuration and does not establish a universal detection bound.

\begin{figure}[!t]
\centering
\includegraphics[width=\columnwidth,keepaspectratio]{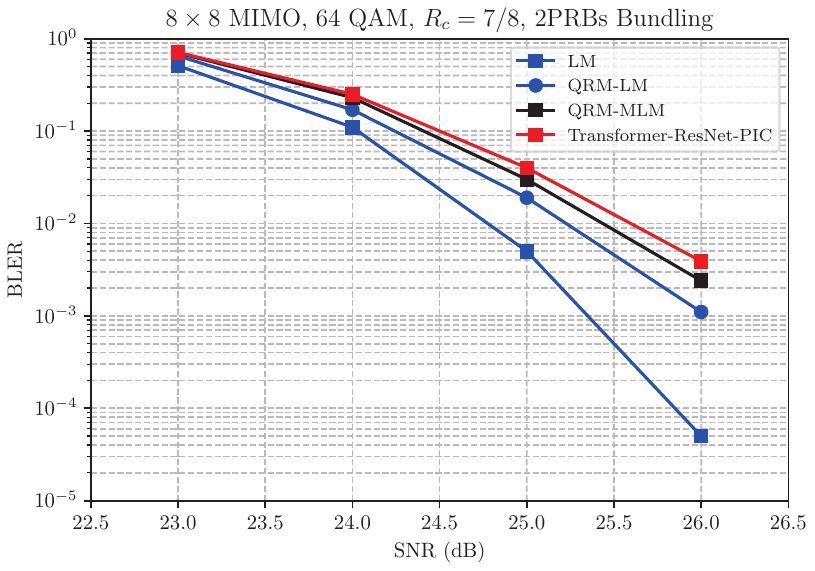}
\caption{Standalone detector BLER with $8\times8$ MIMO, $64$-QAM, $R_c=7/8$, and an ETU channel with $\rho_t=\rho_r=0$. Both detector tests use genie channels and no intercell interference.}
\label{fig:bler_transformer_resnet_pic_8x8mimo_64qam_rc78}
\end{figure}

%
Figure~\ref{fig:bler_transformer_resnet_pic_8x8mimo_64qam_rc78} considers the more demanding $8\times8$ MIMO configuration with $64$-QAM, $R_c=7/8$, and uncorrelated antenna branches under the same interference-free, genie-channel evaluation. Unlike the preceding case, the Transformer-ResNet-PIC detector exhibits a visible gap to LM and slightly underperforms both QRM baselines.

The contrast between the two configurations identifies a limitation of the current detector. With more spatial streams, a larger constellation, and less coding redundancy, its learned soft decisions do not retain the LM-level coded performance observed in Fig.~\ref{fig:bler_transformer_resnet_pic_4x4mimo_16qam_rc13}. Because the antenna dimensions, modulation order, code rate, and channel model all change, these results do not isolate the effect of any one parameter. They nevertheless show that accurate interference reconstruction and channel re-estimation alone cannot ensure LM-level performance when the detection approximation remains limiting. The complete receiver therefore depends on both reliable input estimates and a detector suited to the operating configuration.

\section{Conclusion}

In this article, we proposed an iterative AI receiver for MIMO transmission with finite-alphabet multiuser interference. We specifically considered the practical setting in which covariance-based IRC cannot fully characterize the non-Gaussian, RE-dependent disturbance or recover its instantaneous realization. The proposed receiver couples learned $\bm{Gx}$ reconstruction and cancellation, data-aided desired-channel re-estimation, and Transformer-ResNet-PIC detection through soft feedback. Two receiver architectures and a stage-wise training procedure implement these operations while retaining model-based MMSE-IRC preprocessing. Link-level simulations show that denoising multiuser interference enables the proposed iterative AI receiver to approach the attainable theoretical performance bound. The proposed receiver substantially outperforms the conventional quasi-ML receiver, which relies on LMMSE channel estimation and MMSE-IRC preprocessing. These results confirm that accurate interference reconstruction enables the subsequent channel-estimation and detection gains and constitutes the key requirement for approaching the receiver performance ceiling. Future work will investigate online adaptation to time-varying interference statistics and joint receiver-decoder optimization for additional users and spatial streams.

\appendices

\section{Derivation of the LMMSE Interference Estimate}\label{Appendix_A}
Condition on $\bm{H}$, $\bm{G}$, and the soft state defining $\bar{\bm{s}}$ and $\bm{C}_{\bm{s}}$. The zero-mean, uncorrelated assumptions of Section III-A give
\begin{equation}\label{eqn:app_A_cross_cov}
\begin{split}
\bm{C}_{\bm{z}\bar{\bm{y}}}&\triangleq\mathbf{E}\left\{\bm{z}\bar{\bm{y}}^{H}\right\}\\
&=\mathbf{E}\left\{\bm{G}\bm{x}(\bm{H}\Delta\bm{s}+\bm{G}\bm{x}+\bm{n})^{H}\right\}\\
&=\bm{G}\mathbf{E}\left\{\bm{x}\bm{x}^{H}\right\}\bm{G}^{H}=\bm{G}\bm{G}^{H}.
\end{split}
\end{equation}
and
\begin{equation}\label{eqn:app_A_obs_cov}
\begin{split}
\bm{C}_{\bar{\bm{y}} \bar{\bm{y}}}&\triangleq\mathbf{E}\left\{\bar{\bm{y}}\bar{\bm{y}}^{H}\right\}\\
&=\bm{H}\bm{C}_{\bm{s}}\bm{H}^{H}
+\bm{G}\mathbf{E}\left\{\bm{x}\bm{x}^{H}\right\}\bm{G}^{H}
+\mathbf{E}\left\{\bm{n}\bm{n}^{H}\right\}\\
&=\bm{H}\bm{C}_{\bm{s}}\bm{H}^{H}+\bm{G}\bm{G}^{H}+\sigma^{2}\bm{I}.
\end{split}
\end{equation}
For $\hat{\bm{z}}=\bm{W}_{\bm{z}}^H\bar{\bm{y}}$, expanding the squared-error trace yields
\begin{equation}\label{eqn:app_A_mse_expansion}
\begin{split}
\mathcal{J}\left(\bm{W}_{\bm{z}}^{H}\right)
&\triangleq\mathbf{E}\left\{\left\|\bm{z}-\bm{W}_{\bm{z}}^{H}\bar{\bm{y}}\right\|_{2}^{2}\right\}  \\
&=\operatorname{tr}\left\{\bm{C}_{\bm{z}\bm{z}}\right\}
-2\operatorname{Re}\left\{\operatorname{tr}\left(\bm{W}_{\bm{z}}^{H}\bm{C}_{\bar{\bm{y}}\bm{z}}\right)\right\}  \\
&\quad+\operatorname{tr}(\bm{W}_{\bm{z}}^{H}\bm{C}_{\bar{\bm{y}}\bar{\bm{y}}}\bm{W}_{\bm{z}}).
\end{split}
\end{equation}
where $\bm{C}_{\bm{z}\bm{z}}=\mathbf{E}\{\bm{z}\bm{z}^H\}$ and $\bm{C}_{\bar{\bm{y}}\bm{z}}=\bm{C}_{\bm{z}\bar{\bm{y}}}^H$. Completing the square gives
\begin{equation}
\begin{split}
\mathcal{J}\left(\bm{W}_{\bm{z}}^{H}\right)-\mathcal{J}\left(\left(\bm{W}_{\bm{z}}^{\star}\right)^{H}\right)
=\operatorname{tr}\{&\left(\bm{W}_{z}-\bm{W}_{z}^{\star}\right)^{H}\bm{C}_{\bar{\bm{y}}\bar{\bm{y}}}\\
&\times\left(\bm{W}_{z}-\bm{W}_{z}^{\star}\right)\}\geq 0,
\end{split}
\end{equation}
when the orthogonality condition
\begin{equation}\label{eqn:app_A_normal_eq}
\bm{C}_{\bm{z} \bar{\bm{y}}}-\bm{W}_{\bm{z}}^{H}\bm{C}_{\bar{\bm{y}}\bar{\bm{y}}}=\bm{0}.
\end{equation}
holds. Since $\sigma^2>0$, the observation covariance is positive definite, so the unique solution is
\begin{equation}\label{eqn:app_A_filter}
\begin{split}
\bm{W}_{z}^{H}
&=\bm{C}_{\bm{z}\bar{\bm{y}}}\bm{C}_{\bar{\bm{y}}\bar{\bm{y}}}^{-1}\\
&=\bm{G}\bm{G}^{H}
\left(\bm{G}\bm{G}^{H}+\bm{H}\bm{C}_{\bm{s}}\bm{H}^{H}
+\sigma^{2}\bm{I}\right)^{-1}.
\end{split}
\end{equation}
Multiplication by $\bar{\bm{y}}$ gives \eqref{eqn:Gx_est}. Similarly, $\hat{\bm{x}}=\bm{G}^H\bm{C}_{\bar{\bm{y}}\bar{\bm{y}}}^{-1}\bar{\bm{y}}$ and $\hat{\bm{z}}=\bm{G}\hat{\bm{x}}$. This second-order derivation establishes linear optimality; finite-alphabet Bayesian MMSE estimation is generally nonlinear.

\section{Mathematical Foundation of Transformer-ResNet-PIC}\label{Appendix_B}
Consider the interference-cancelled model
\begin{equation}\label{eqn:mimo_detection}
  \bm{y} = \bm{Hs} + \tilde{\bm{n}},
\end{equation}
with Gaussian residual $\tilde{\bm{n}}\sim\mathcal{CN}(\bm{0},\bm{R})$; perfect cancellation gives $\bm{R}=\sigma^2\bm{I}$. For constellation $\mathcal{A}$, $|\mathcal{A}|=M$, MAP detection under zero-one loss is
\begin{equation}\label{eqn:map_detection}
  \begin{split}
     \hat{\bm{s}}_{\mathrm{MAP}}
    &=\argmax\limits_{\bm{s}\in\mathcal{A}^{N_t}}p(\bm{s}|\bm{y}) \\
    &=\argmax\limits_{\bm{s}\in\mathcal{A}^{N_t}}
    \frac{p(\bm{y}|\bm{s})p(\bm{s})}
    {\sum\limits_{\tilde{\bm{s}}\in\mathcal{A}^{N_t}}p(\bm{y}|\tilde{\bm{s}})p(\tilde{\bm{s}})},
  \end{split}
\end{equation}
The candidate-independent denominator gives the equivalent log objective
\begin{equation}\label{eqn:log_map_detection}
  \hat{\bm{s}}_{\mathrm{MAP}}
  \propto \argmax\limits_{\bm{s}\in\mathcal{A}^{N_t}}
  \left\{\log p(\bm{y}|\bm{s})+\log p(\bm{s})\right\}.
\end{equation}
Using the Gaussian likelihood
\begin{equation}\label{eqn:complex_likelihood}
  p(\bm{y}|\bm{s})
  =\frac{1}{\pi^{N_r}\det(\bm{R})}
  \exp\left\{-(\bm{y}-\bm{Hs})^{H}\bm{R}^{-1}(\bm{y}-\bm{Hs})\right\}.
\end{equation}
gives
\begin{equation}\label{eqn:discrete_map_metric}
  \hat{\bm{s}}_{\mathrm{MAP}}
  =\argminA\limits_{\bm{s}\in\mathcal{A}^{N_t}}
  \left\{(\bm{y}-\bm{Hs})^{H}\bm{R}^{-1}(\bm{y}-\bm{Hs})-\log p(\bm{s})\right\}.
\end{equation}
Evaluating $M^{N_t}$ hypotheses motivates an iterative approximation.

For a continuous variable $\bm{v}\in\mathbb{C}^{N_t\times1}$ and differentiable surrogate prior $p_\theta$, define
\begin{equation}\label{eqn:relaxed_log_posterior}
  \mathcal{J}_{\theta}(\bm{v})
  \triangleq -(\bm{y}-\bm{Hv})^{H}\bm{R}^{-1}(\bm{y}-\bm{Hv})
  +\log p_{\theta}(\bm{v}).
\end{equation}
The first-order complex expansion at $\bm{v}^{(\ell)}$ is
\begin{equation}
\begin{split}
  \mathcal{J}_{\theta}\left(\bm{v}^{(\ell)}+\Delta\bm{v}\right)
  ={}&\mathcal{J}_{\theta}\left(\bm{v}^{(\ell)}\right)\\
  &+2\operatorname{Re}\!\left\{
  \left(\nabla_{\bm{v}^{*}}\mathcal{J}_{\theta}(\bm{v}^{(\ell)})\right)^{H}
  \Delta\bm{v}\right\}\\
  &+o(\|\Delta\bm{v}\|_{2}),
\end{split}
\end{equation}
where $\nabla_{\bm{v}^*}$ is the Wirtinger gradient. Regularizing the unbounded linear model gives
\begin{equation}
\begin{split}
  \mathcal{Q}^{(\ell)}(\Delta\bm{v})\triangleq {}&
  2\operatorname{Re}\!\left\{
  \left(\nabla_{\bm{v}^{*}}\mathcal{J}_{\theta}(\bm{v}^{(\ell)})\right)^{H}\Delta\bm{v}
  \right\}\\
  &-\frac{1}{\gamma_{\ell}}\|\Delta\bm{v}\|_{2}^{2},
\end{split}
\end{equation}
Here $\gamma_\ell>0$ is an algorithmic step-size parameter, not an additional prior; its real-embedding proximal step size is $\gamma_\ell/2$. The update solves
\begin{equation}\label{eqn:local_gradient_subproblem}
\begin{split}
  \Delta\bm{v}^{(\ell)}=\argmax\limits_{\Delta\bm{v}}\ \Big\{&
  2\operatorname{Re}\!\left\{
  \left(\nabla_{\bm{v}^{*}}\mathcal{J}_{\theta}\left(\bm{v}^{(\ell)}\right)\right)^{H}\Delta\bm{v}
  \right\}\\
  &-\frac{1}{\gamma_{\ell}}\|\Delta\bm{v}\|_{2}^{2}\Big\},
\end{split}
\end{equation}
with stationarity condition
\begin{equation}
  \nabla_{\Delta\bm{v}^{*}}\mathcal{Q}^{(\ell)}
  =\nabla_{\bm{v}^{*}}\mathcal{J}_{\theta}\left(\bm{v}^{(\ell)}\right)
  -\frac{1}{\gamma_{\ell}}\Delta\bm{v}=\bm{0}.
\end{equation}
Strict concavity gives $\Delta\bm{v}^{(\ell)}=\gamma_\ell\nabla_{\bm{v}^*}\mathcal{J}_\theta(\bm{v}^{(\ell)})$ and
\begin{equation}\label{eqn:map_gradient_iteration}
\begin{split}
  \bm{v}^{(\ell+1)}
  &=\bm{v}^{(\ell)}+\gamma_{\ell}\nabla_{\bm{v}^{*}}
  \mathcal{J}_{\theta}(\bm{v}^{(\ell)})\\
  &=\bm{v}^{(\ell)}+\gamma_{\ell}\Big[
  \bm{H}^{H}\bm{R}^{-1}\left(\bm{y}-\bm{H}\bm{v}^{(\ell)}\right)\\
  &\hspace{24mm}
  +\nabla_{\bm{v}^{*}}\log p_{\theta}\left(\bm{v}^{(\ell)}\right)\Big].
\end{split}
\end{equation}
For an $L$-Lipschitz real-embedding gradient, the lower smoothness bound is
\begin{equation}\label{eqn:Lipschitz}
\begin{split}
  \mathcal{J}_{\theta}\left(\bm{v}+\Delta\bm{v}\right)
  \geq {}&\mathcal{J}_{\theta}(\bm{v})
  +\mathrm{Re}\!\left\{
  \left(\nabla_{\bm{v}^{*}}\mathcal{J}_{\theta}(\bm{v})\right)^{H}
  \Delta\bm{v}\right\}\\
  &-\frac{L}{2}\left\|\Delta\bm{v}\right\|_{2}^{2}.
\end{split}
\end{equation}
When $0<\gamma_\ell\leq2/L$, the local quadratic model lower-bounds the objective increment. Since $\mathcal{Q}^{(\ell)}(\bm{0})=0$,
\begin{equation}
  \begin{split}
     \mathcal{J}_{\theta}\left(\bm{v}^{\ell+1}\right) & \geq  \mathcal{J}_{\theta}\left(\bm{v}^{\ell}\right) + \mathcal{Q}^{(\ell)}\left(\Delta \bm{v}^{(\ell)} \right)\\
       & = \mathcal{J}_{\theta}\left(\bm{v}^{(\ell)}\right)^{(\ell)} + \gamma_{\ell}\left\|\nabla_{\bm{v}}^{*}\mathcal{J}_{\theta}\left(\bm{v}^{(\ell)}\right)\right\|_{2}^{2}.
  \end{split}
\end{equation}
The objective therefore increases; boundedness and step sizes bounded away from zero imply a vanishing gradient norm. Nonconcavity precludes a global-MAP guarantee. The update combines observation consistency and a surrogate prior score.

Suppressing $i$, define the IRC statistics
\begin{subequations}\label{eqn:irc_statistics}
\begin{align}
  \bm{H}_{irc} &\triangleq \bm{H}^{H}\left(\bm{H}\bm{H}^{H}+\bm{R}\right)^{-1}\bm{H},\\
  \bm{s}_{irc} &\triangleq \bm{H}^{H}\left(\bm{H}\bm{H}^{H}+\bm{R}\right)^{-1}\bm{y},\\
  \bm{K} &\triangleq \bm{H}^{H}\bm{R}^{-1}\bm{H}, \\
  \bm{b} &\triangleq \bm{H}^{H}\bm{R}^{-1}\bm{y}.
\end{align}
\end{subequations}
The matrix inversion lemma gives
\begin{subequations}\label{eqn:irc_sufficient_relation}
\begin{align}
  \bm{H}_{irc} &=(\bm{I}+\bm{K})^{-1}\bm{K},\\
  \bm{s}_{irc} &=(\bm{I}+\bm{K})^{-1}\bm{b},\\
  \bm{I}-\bm{H}_{irc} &=(\bm{I}+\bm{K})^{-1}.
\end{align}
\end{subequations}
and the likelihood-gradient identity
\begin{equation}\label{eqn:gradient_from_irc}
  \bm{H}^{H}\bm{R}^{-1}\left(\bm{y} - \bm{Hv}\right) = \left(\bm{I} - \bm{H}_{irc}\right)^{-1}\left(\bm{s}_{irc}-\bm{H}_{irc}\bm{v}\right).
\end{equation}
Thus, $\bm{s}_{irc}$ and $\bm{H}_{irc}$ encode the linear estimate and stream coupling, while $\tilde{\bm{y}}=\bm{s}_{irc}-\bm{H}_{irc}\bm{v}$ is a preconditioned data-consistency residual.

The learned detector replaces prescribed scores and step sizes by
\begin{subequations}\label{eqn:learned_inference_update}
\begin{align}
    \bm{q}^{(\ell)} &= \mathcal{T}_{\phi_{\ell}}\left(\bm{H}_{irc}, \tilde{\bm{y}}^{(\ell-1)}, \bm{P}_{s}^{(\ell-1)}, \bm{q}^{(\ell-1)}\right), \\
    \bm{L}_{\bm{s}}^{(\ell)} &= \mathcal{D}_{\phi_{\ell}}\left(\bm{q}^{(\ell)}\right), \\
    \bm{P}_{\bm{s}}^{(\ell)} &= \mathrm{softmax}\left(\bm{L}_{\bm{s}}^{(\ell)}\right), \\
    \bm{s}^{(\ell)} &= \sum\limits_{m=1}^{M}\alpha_{m}\bm{P}_{\bm{s}}^{(\ell)}\left[:,:,m \right].
\end{align}
\end{subequations}
Here $\bm{q}^{(\ell)}$ is the latent state, and $\mathcal{T}_{\phi_\ell}\left(\cdot\right)$ and $\mathcal{D}_{\phi_\ell}\left(\cdot\right)$ are the Transformer and MLP. Softmax and posterior-mean readout coincide with \eqref{eqn:prob_softmax} and \eqref{eqn:soft_decision}. The residual follows \eqref{eqn:pic}. The block can represent a preconditioned MAP-gradient update, but this structural interpretation gives no general convergence guarantee unless the learned direction enforces ascent of \eqref{eqn:relaxed_log_posterior}.


\bibliographystyle{IEEEtran}

\end{document}